\documentclass[10pt,preprint]{aastex}

\pdfoutput=1

\usepackage{url}
\usepackage{grffile}

\newcounter{column_number}
\newcommand{\numberthecolumn}{\colhead{(\arabic{column_number})}\stepcounter{column_number}}

\newcommand{\hii}{H{\scriptsize II} }
\newcommand{\etacar}{$\eta$~Car }
\newcommand{\Chandra}{{\em Chandra} }

\shorttitle{Chandra Carina Complex Project}
\shortauthors{Townsley et al.} 

\begin{document}

\title{THE INTEGRATED DIFFUSE X-RAY EMISSION OF THE CARINA NEBULA COMPARED TO OTHER MASSIVE STAR-FORMING REGIONS}

\author{ Leisa K. Townsley\altaffilmark{*}\altaffilmark{1}, 
Patrick S. Broos\altaffilmark{1},
You-Hua Chu\altaffilmark{2},
Robert A. Gruendl\altaffilmark{2},
M. S. Oey\altaffilmark{3},
Julian M. Pittard\altaffilmark{4}
}

\altaffiltext{*}{townsley@astro.psu.edu} 

\altaffiltext{1}{Department of Astronomy \& Astrophysics, 525 Davey Laboratory, Pennsylvania State University, University Park, PA 16802, USA}

\altaffiltext{2}{Department of Astronomy, University of Illinois at Urbana-Champaign, 1002 West Green Street, Urbana, IL 61801, USA}

\altaffiltext{3} {Department of Astronomy, University of Michigan, 830 Dennison Building, Ann Arbor, MI   48109-1042, USA}

\altaffiltext{4} {School of Physics and Astronomy, The University of Leeds, Woodhouse Lane, Leeds LS2 9JT, UK}


\begin{abstract}
The \Chandra Carina Complex Project (CCCP) has shown that the Carina Nebula displays bright, spatially-complex soft diffuse X-ray emission.  Here we `sum up' the CCCP diffuse emission work by comparing the global morphology and spectrum of Carina's diffuse X-ray emission to other famous sites of massive star formation with pronounced diffuse X-ray emission:  M17, NGC~3576, NGC~3603, and 30~Doradus.  All spectral models require at least two diffuse thermal plasma components to achieve adequate spectral fits, a softer component with kT = 0.2--0.6~keV and a harder component with kT = 0.5--0.9~keV.  In several cases these hot plasmas appear to be in a state of non-equilibrium ionization that may indicate recent and current strong shocks.  A cavity north of the embedded giant \hii region NGC~3576 is the only region studied here that exhibits hard diffuse X-ray emission; this emission appears to be nonthermal and is likely due to a recent cavity supernova, as evidenced by a previously-known pulsar and newly-discovered pulsar wind nebula also seen in this cavity.  All of these targets exhibit X-ray emission lines that are not well-modeled by variable-abundance thermal plasmas and that might be attributed to charge exchange at the shock between the hot, tenuous, X-ray-emitting plasma and cold, dense molecular material; this is likely evidence for dust destruction at the many hot/cold interfaces that characterize massive star-forming regions. 

\end{abstract}

\keywords{HII regions --- X-rays: individual (Carina, M17, NGC~3576, NGC~3603, 30~Doradus)}


\section{INTRODUCTION \label{sec:intro}}

This is the final paper in the {\em Special Issue} devoted to the \Chandra Carina Complex Project (CCCP), a 1.2-Ms, 1.42 square degree mosaic of the Great Nebula in Carina obtained with the Imaging Array of the Advanced CCD Imaging Spectrometer camera \citep[ACIS-I,][]{Garmire03} on the {\em Chandra X-ray Observatory}.  An introduction and overview of the CCCP is provided by \citet{Townsley11a}; 14 subsequent papers in this {\em Special Issue} provide in-depth studies of the constituents of the Great Nebula in Carina.  In particular, the 15th paper in the {\em Special Issue} gives a detailed spatio-spectral analysis of the diffuse X-ray emission in Carina \citep{Townsley11b}; that work details the development of, and explains the need for, a complicated spectral model of Carina's diffuse X-ray emission. It gives the foundations for our argument that a newly-discovered X-ray emission mechanism in \hii regions, possibly charge exchange between the cold neutral material and the hot X-ray-emitting plasma \citep{Lallement04,Ranalli08}, may be operating in Carina.  Our primary goal in this paper is to assess whether this X-ray emission mechanism---whether it turns out to be charge exchange or some other previously-unrecognized physical process---is unique to Carina or ubiquitous in \hii regions with bright diffuse X-ray emission.  We encourage the reader to review the CCCP diffuse emission paper; some familiarity with that effort is assumed here.

Separating truly diffuse X-ray emission due to hot plasmas in \hii regions from the collective X-ray emission of many hundreds to thousands of unresolved pre-Main Sequence (pre-MS) stars that occupy the same space has proved difficult over the years, even with the high spatial resolution, good sensitivity, and wide field coverage of \Chandra and ACIS-I.  Early efforts were summarized by \citet{Townsley03}.  Some more recent examples include studies of NGC~6334 \citep{Ezoe06}, RCW~38 \citep{Wolk06}, Westerlund~1 \citep{Muno06}, and W40 \citep{Kuhn10}.

In order to gain a broader understanding of the CCCP results, it is instructive to consider Carina's X-ray properties in the context of those of other giant \hii regions (GHIIRs) in the Milky Way and other nearby galaxies.  Carina's X-ray emission could be compared to that of other regions in a variety of ways, e.g., massive star X-ray luminosities and spectra, X-ray luminosity functions of pre-MS stars, lightcurve characteristics for X-ray-variable point sources, or diffuse X-ray emission properties.  Here we choose to concentrate on the last of these examples, comparing Carina's integrated diffuse X-ray emission to that of some other famous massive star-forming regions studied by {\em Chandra}.  

In the introduction to the CCCP \citep[][Section~5.7]{Townsley11a}, we presented some simple scalings to place the CCCP in the context of the \Chandra Orion Ultradeep Project \citep[COUP,][]{Feigelson05,Getman05}, a more familiar X-ray survey of the Orion Nebula Cluster (ONC).  \citet{Gudel08} detected diffuse X-ray emission near the ONC with {\em XMM-Newton}; its thermal plasma temperature (kT$<$0.2~keV) is softer than the plasma in Carina \citep{Townsley11b}, with an order of magnitude lower intrinsic surface brightness.  Here we concentrate on the integrated emission from some other GHIIRs that, like Carina, display bright diffuse X-ray emission:  M17, NGC~3576, NGC~3603, and 30~Doradus \citep{Townsley03,Townsley06a,Townsley09a,Townsley09b}.  We examine the global morphology and intrinsic spectral properties of this emission to decide whether Carina's integrated diffuse X-ray emission properties should be considered typical or extreme.

We emphasize that our goal here is a brief preliminary study of global, integrated properties of the diffuse X-ray emission in a variety of massive star-forming regions, chosen because they exhibit bright unresolved X-ray emission in their \Chandra data, not because they are typical examples of \hii region diffuse X-ray emission (which is usually quite faint and more comparable to the ONC).  Detailed studies of the diffuse X-ray emission in the targets presented here, with spatially-resolved spectroscopy and derived parameter maps analogous to the CCCP diffuse emission study \citep{Townsley11b}, will be presented in separate future papers.  Again our motivation for studying the global diffuse X-ray emission properties of these famous GHIIRs is simply to get a sense for the degree of similarity or diversity that such complexes might display, in order to further our understanding of the physical processes at work there.

Before comparing the diffuse X-ray emission in these GHIIRs to the CCCP results for Carina, we must take time to introduce each target and its \Chandra data (Section~\ref{sec:observations}).  The spectral analysis for this diffuse emission is then presented, again ordered by individual target, in Section~\ref{sec:spectra}, using a complicated spectral model based on the model developed by \citet{Townsley11b} for Carina.  The details and justification for the Carina X-ray spectral model can be found in that paper and we encourage readers to look there for an in-depth explanation of the model's form; here we simply test its efficacy on other targets.  In Section~\ref{sec:discussion} we summarize the global diffuse X-ray emission properties of each target and compare each in turn to Carina.  We summarize the findings of this initial, cursory study in Section~\ref{sec:conclusions}.  Again we emphasize that more careful, spatially-resolved X-ray spectral analysis for each of these targets (and for many more, less extreme \hii regions) will be the subject of future work.  The complexity found in the CCCP diffuse emission study of Carina suggests that similar richness---and perhaps more surprises---will be found in these future detailed studies of other GHIIRs with bright diffuse X-ray emission.








\section{{\em CHANDRA} IMAGING OF BRIGHT DIFFUSE X-RAY EMISSION IN MASSIVE STAR-FORMING REGIONS \label{sec:observations}}

The first step in the quantitative analysis of diffuse X-ray emission in star-forming complexes is the identification and removal of X-ray point sources.  For {\em Chandra}/ACIS observations, this painstaking process is described in \citet{Broos10}; its application to the Carina Nebula---which resulted in a census of $>$14,000 X-ray point sources---is described in the CCCP catalog paper \citep{Broos11}.  The same methods were used for point source identification and extraction for the other targets presented here; catalogs of those point sources will be presented in future target-specific papers.

Pre-MS stars are of course ubiquitous in young star-forming regions; they are bright X-ray emitters, with a typical spectrum consisting of a soft thermal plasma component with $kT \sim 0.86$~keV and a brighter, harder thermal component with $kT \sim 2.6$~keV \citep{Preibisch05}.  These objects are of particular concern for diffuse emission studies because of their soft X-ray component; large unresolved pre-MS populations in short \Chandra observations or in X-ray observations using telescopes with larger point spread functions can masquerade as soft diffuse emission.  In the CCCP diffuse emission paper \citep{Townsley11b}, we identified a thermal plasma component in Carina's ``diffuse'' spectrum that is likely due to the harder emission from unresolved pre-MS stars and used its X-ray luminosity to predict the fraction of Carina's soft diffuse emission that was due to contamination from pre-MS stars.  We found only $\sim$3\% contamination there, but lower-sensitivity studies likely have higher contamination from unresolved pre-MS stars, as we will see below.

In order to study the morphology of diffuse X-ray emission in massive star-forming regions, we developed an adaptive-kernel smoothing technique that works with images where X-ray point sources have been masked \citep{Townsley03,Broos10}.  In the images that follow, we present these smoothed masked images with the apertures used for point source extraction \citep[e.g.,][]{Broos11} overlaid; this shows the spatial distribution of detected X-ray point sources but prevents resolved point source emission from being confused with the smoothed diffuse emission.  The locations of excised point sources are shown so that the reader has some idea of the spatial distribution of detected point sources.  Due to the varying point spread function and vignetting in the \Chandra mirrors, the point source detection sensitivity is not uniform across the ACIS-I field of view; this issue is explored in detail for the complicated CCCP mosaic of Carina in \citet{Broos11} and \citet{Feigelson11}.

When foreground absorbing columns are low enough and/or our observations are sensitive enough to provide adequate counts, we present images in three narrow, soft X-ray bands (defined below) that are good for highlighting typically soft diffuse features; this is the case for Carina, M17, and 30~Doradus.  For NGC~3576 and NGC~3603, short observations and high absorbing columns give so few soft X-ray counts that no meaningful structure is apparent in the soft narrow-band smoothed images, so we present images in the more traditional, broader soft X-ray band of 0.5--2~keV and in the hard band (2--7~keV), supplemented with mid-infrared (mid-IR) {\em MSX} \citep[Midcourse Space Experiment,][]{Price01} and {\em Spitzer} images to provide context.

\subsection{Carina}

\begin{figure}[htb] 
\begin{center}  
\includegraphics[width=1.0\textwidth]{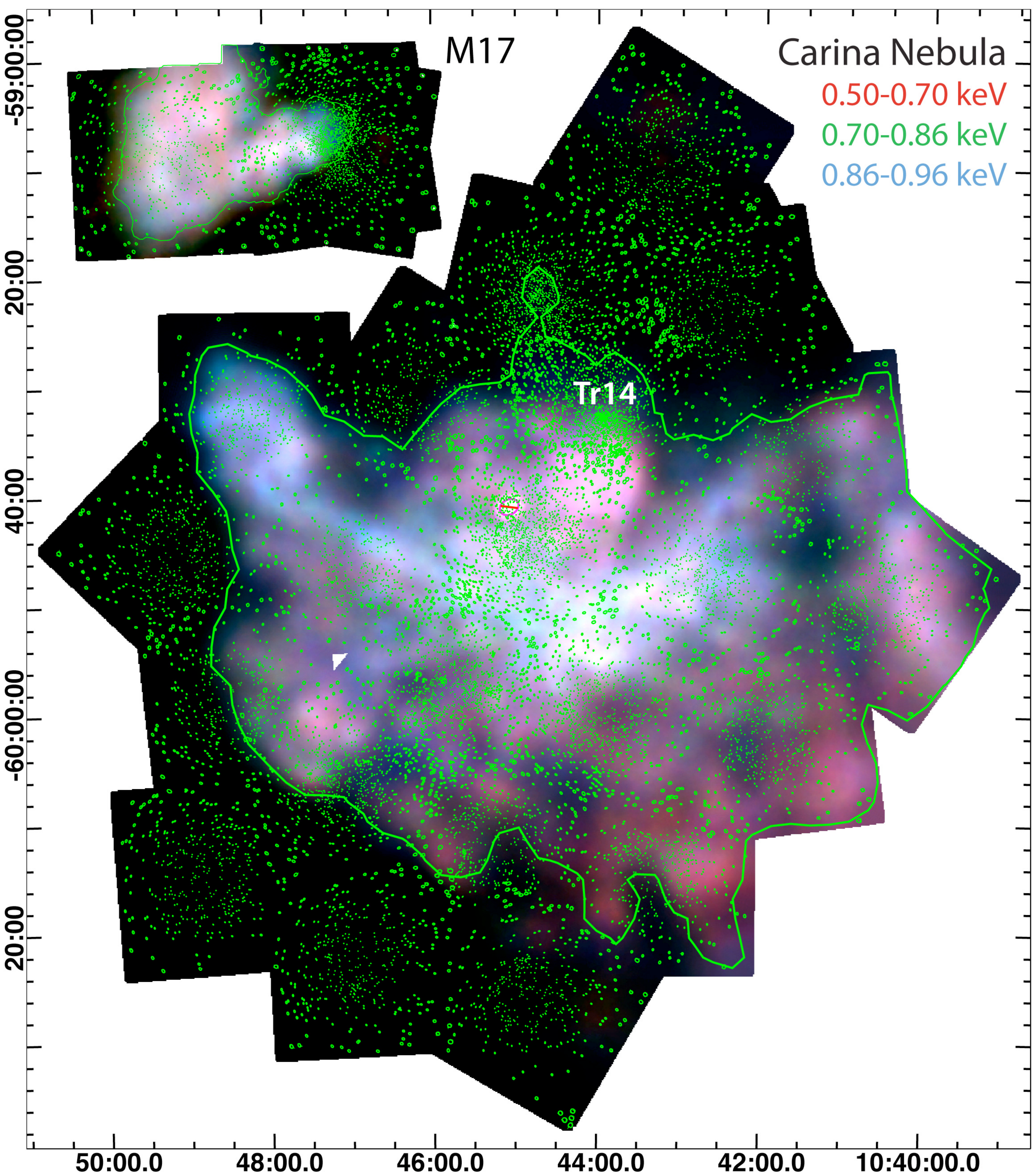}
\caption{Soft diffuse emission in the Carina Nebula and a comparison to M17.  Here and throughout this paper, image coordinates are celestial J2000.  For each target, the diffuse regions extracted for the ``global'' diffuse spectra are outlined by heavy green lines.  The red oval with a line through it northeast of the CCCP field center shows the region around \etacar that was excised from Carina's integrated diffuse spectrum.  Small green polygons show the locations of X-ray point sources from the CCCP that were excised to create this image.  For comparison, the ACIS-I mosaic of M17 is shown as an inset in the upper left; it shows similar soft-band images and ACIS point source extraction regions and is roughly to scale, since M17 and Carina are at similar distances.
} 
\label{fig:carina+m17image}
\end{center}
\end{figure}

The Carina star-forming complex (D$\sim$2.3~kpc) is a ``cluster of clusters'' containing 8 open clusters with at least 66 O stars, 3 Wolf-Rayet stars, and the luminous blue variable \etacar \citep{Smith08}.  Figure~\ref{fig:carina+m17image} shows the soft diffuse X-ray emission in the Carina Nebula.  This is the same adaptively-smoothed data used in the CCCP diffuse emission paper \citep[][Figure~1]{Townsley11b}, scaled slightly differently here.  The $>$14,000 X-ray point sources detected by the CCCP have been excised in the image but replaced with their extraction regions (shown as green polygons that trace the size of the point spread function).  

Detailed spatially-resolved spectral fitting, parameter maps, and inferred physical properties of the diffuse plasma in Carina were presented in \citet{Townsley11b}.  Multiwavelength images placing Carina's diffuse X-ray emission in context were presented in that paper and in the CCCP introductory paper \citep{Townsley11a}.  To summarize, Carina's diffuse X-ray emission is soft, with multiple plasma components that exhibit a range of temperatures and absorptions.  Not all plasma components are in ionization equilibrium, although most are.  The plasma appears to be filling cavities in the Nebula that are outlined by dense ionized gas and/or IR emission from PAH's and heated dust.  The brightest diffuse emission is not centered on the densest stellar clusters, nor does it closely trace the distribution of massive stars in the complex.  Based on unmodeled lines seen in the X-ray spectrum, \citet{Townsley11b} suggest that charge exchange emission \citep{Lallement04} contributes substantially to the X-ray spectrum in many parts of Carina, revealing dust destruction at the interfaces between $\sim 10^{7}$~K plasma and $\sim 10^{1-2}$~K molecular material, including molecular clouds, dust pillars, and cold fragments of neutral material remaining in the region.  

\clearpage

\subsection{M17}

\begin{figure}[htb] 
\begin{center}  
\includegraphics[width=1.0\textwidth]{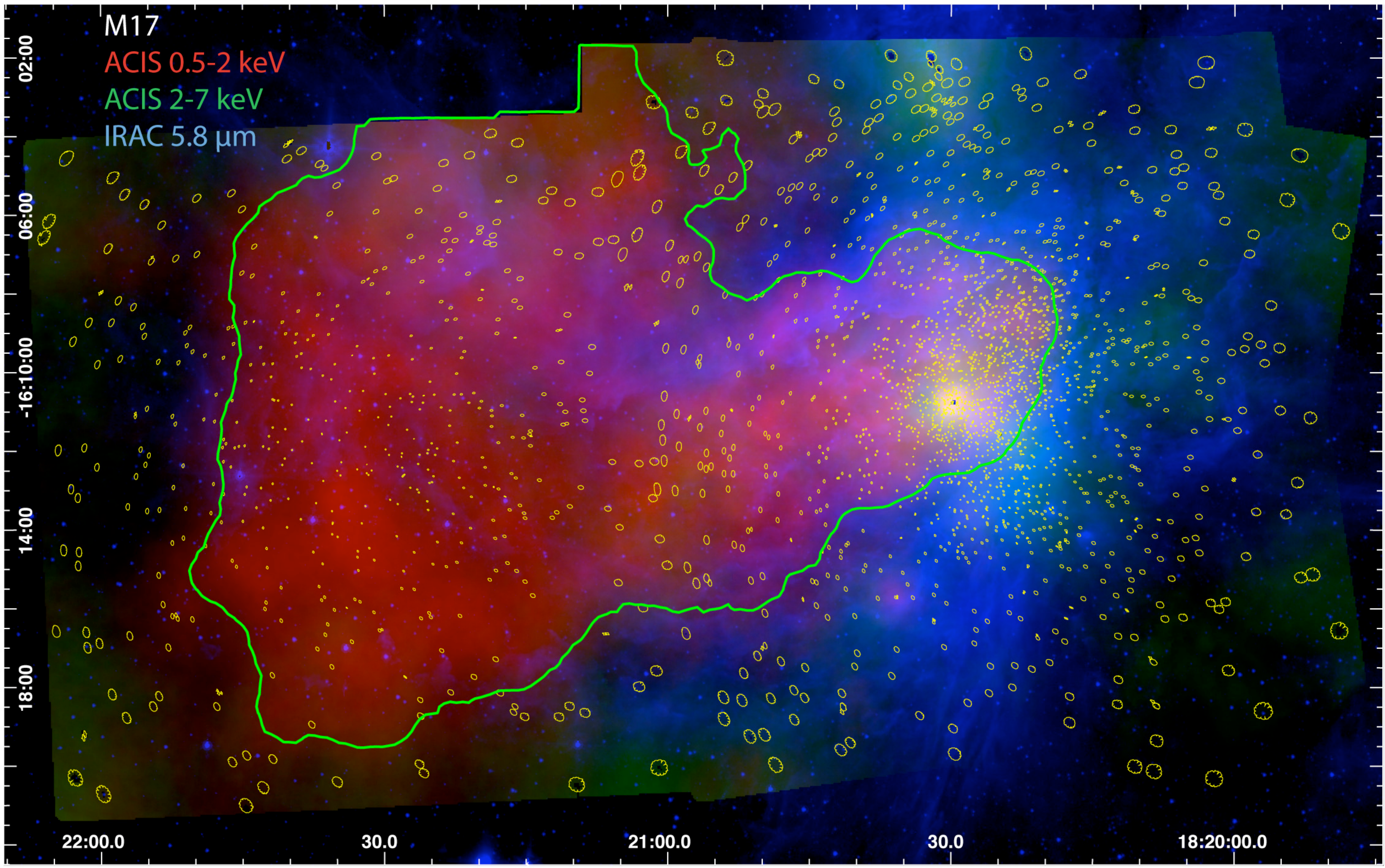}
\caption{Diffuse X-ray emission in M17.  The ACIS-I diffuse mosaic (point sources excised then images adaptively smoothed) is shown in soft-band (0.5--2~keV) and hard-band (2--7~keV) superposed on the GLIMPSE mosaic of 5.8~$\mu$m mid-IR emission from {\em Spitzer}/IRAC \citep{Churchwell09}.  The extraction region for global diffuse X-ray emission is shown in green.  Extraction regions for the $\sim$2700 X-ray point sources that were removed from this mosaic are shown in yellow. 
} 
\label{fig:m17image}
\end{center}
\end{figure}

M17 is one of the closest (D=2.1~kpc), brightest, and youngest (age$\sim$0.5~Myr) GHIIRs known \citep{Chini08,Hoffmeister08}.  Its ionizing cluster, NGC~6618, suffers an average extinction of $A_{V} = 8$~mag \citep{Hanson97}.  An early, 40-ks \Chandra image of NGC~6618 was studied by \citet{Townsley03}, who mapped the diffuse X-ray emission, and \citet{Broos07}, who catalogued almost 900 X-ray sources in the field, including X-ray detections of the 14 known O stars.  A subsequent, longer \Chandra observation maps NGC~6618 and the diffuse X-ray emission as it emerges from M17's molecular cloud on the east side of NGC~6618 (Figure~\ref{fig:m17image}), finding $\sim$700 point sources in this eastern field and $\sim$2000 sources in and around NGC~6618 \citep{Townsley09a}.  

Figure~\ref{fig:carina+m17image} shows an adaptively-smoothed image of M17, prepared in the same way as the Carina image shown in that figure, and displayed as an inset.  The M17 {\em Chandra}/ACIS mosaic is shown there at roughly the same scale as the CCCP mosaic; since the two star-forming regions are at roughly the same distance, we do not need to rebin the M17 image to match the CCCP resolution.  The extraction region used for M17's diffuse X-ray emission is shown in green; it was obtained by contouring the apparent surface brightness in a smoothed soft-band (0.5--2~keV) image.  Note that the NGC~6618 pointing in the M17 mosaic is quite deep, with a total ACIS-I integration time of 320~ks; the eastern pointing is shallower with 92~ks, but this is still deeper than most of the CCCP mosaic, which had a nominal exposure time of 60~ks per pointing \citep{Townsley11a}.

M17's ionizing cluster, NGC~6618, most closely resembles the Carina cluster Trumpler~14 (Tr14) in morphology and age \citep[e.g.,][]{Sana10}.  It is strongly centrally-concentrated, with most of its O stars in its central arcminute.  Its diffuse X-ray emission appears to be channeled towards the edge of the M17 molecular cloud, emerging from a narrow ``crevice'' flanked by the famous northern and southern ionization bars into a larger cavity to the east \citep{Townsley03,Povich07}.  Townsley et al.\ noted that the apparent surface brightness of M17's diffuse X-ray emission is quite high compared to many other famous Galactic massive star-forming regions (e.g., the Rosette, Lagoon, and Eagle Nebulae).

Comparing M17's soft diffuse X-ray emission to that of Carina in Figure~\ref{fig:carina+m17image}, M17 most resembles the region southeast of Tr14 and west of \etacar (centered at RA$\sim10^{h}44^{m}$, Dec$\sim-59^{\circ}38\arcmin$) that shows bright, soft diffuse emission sharply cut off to the west by the Carina I molecular cloud \citep{Townsley11b}.  If this analogy is correct, perhaps that part of Carina's diffuse emission is due to Tr14's massive stellar winds, as we believe is the case for M17's diffuse X-ray emission \citep{Townsley03}.  The CCCP diffuse emission paper \citep[][Figure~13b]{Townsley11b} shows that the diffuse emission immediately surrounding Tr14 has high intrinsic surface brightness, even though its apparent surface brightness is low due to obscuration, indicating that the bright diffuse X-ray emission southeast of Tr14 is probably an extension of this cluster's hot plasma.

\subsection{NGC~3576}

\begin{figure}[htb] 
\begin{center}  
\includegraphics[width=0.49\textwidth]{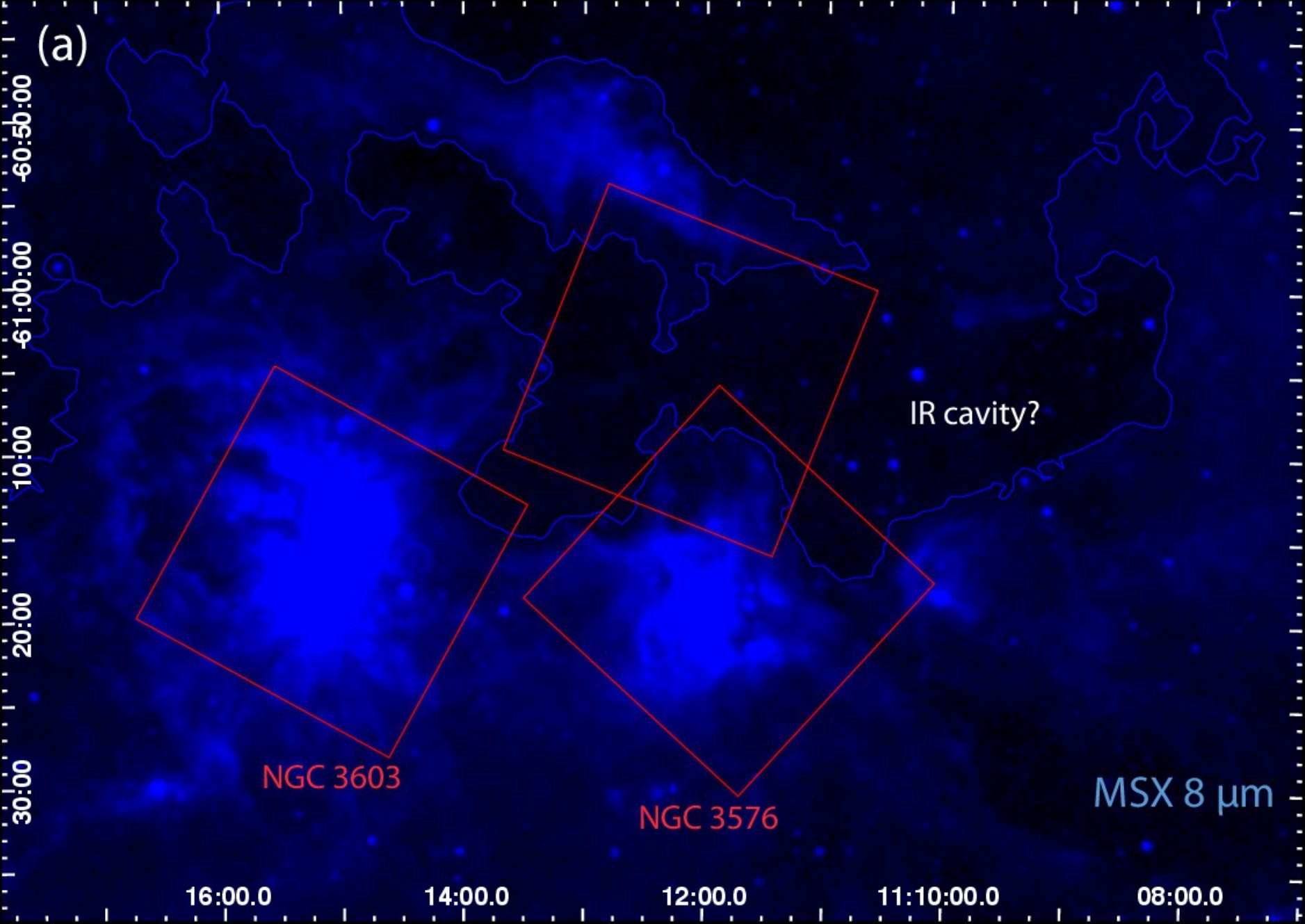}
\includegraphics[width=0.49\textwidth]{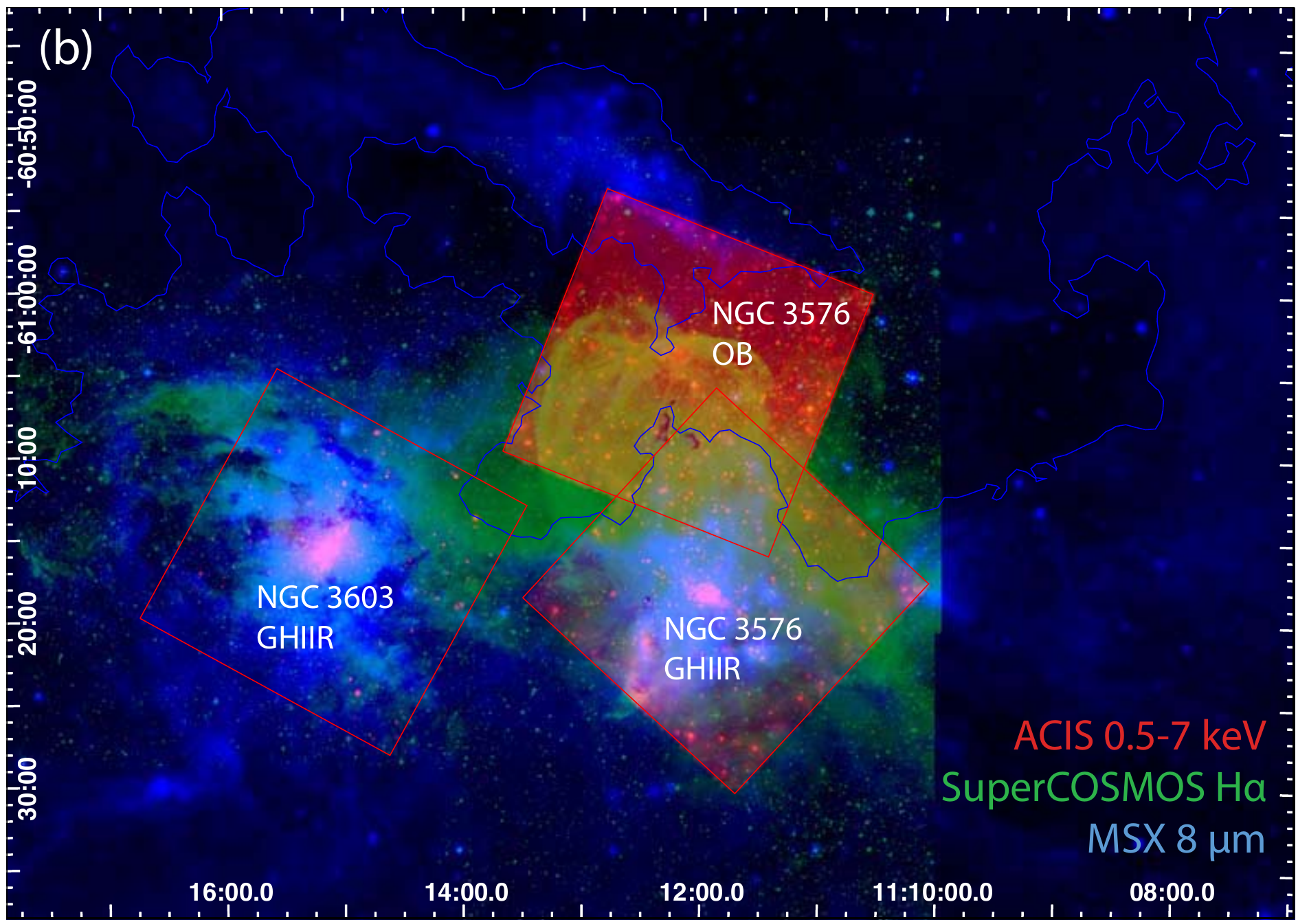}\\
\vspace*{0.1in}
\includegraphics[width=0.8\textwidth]{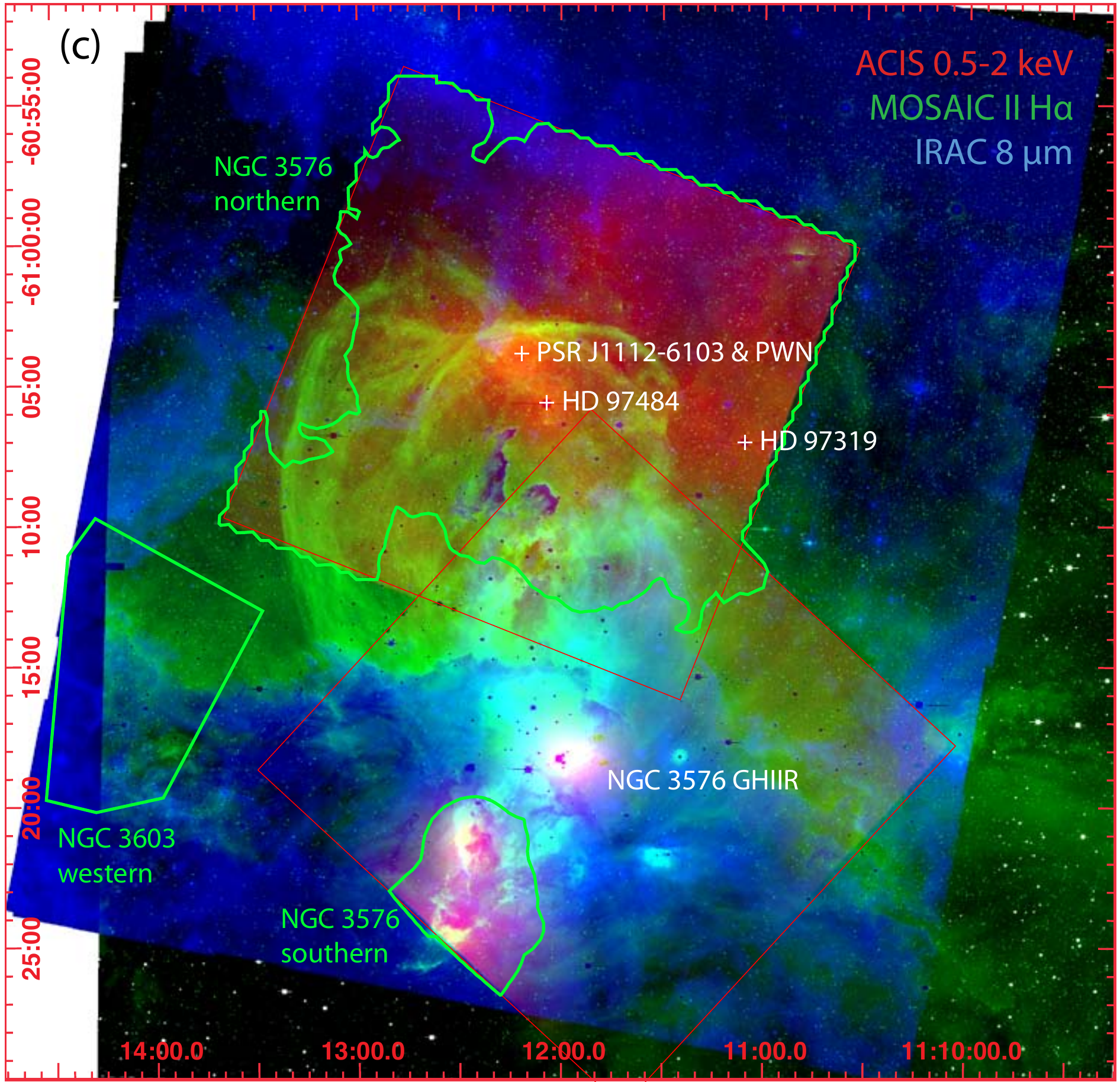}
\caption{The neighborhood of NGC~3576.
(a)  An {\em MSX} image of NGC~3576, NGC~3603, and their surroundings.  ACIS-I pointings on these targets are outlined (in red); a contour of 8~$\mu$m surface brightness (in blue) highlights a possible cavity in the IR ISM.
(b)  The same {\em MSX} image as in (a), now including the SuperCOSMOS H$\alpha$ image \citep{Parker05} and total-band (0.5--7~keV) ACIS-I images (with the X-ray point sources still in place) made using {\it csmooth} \citep{Ebeling02}.
(c)  A new depiction of NGC~3576's ISM:  ACIS-I soft-band (0.5--2~keV) diffuse X-ray emission (point sources removed, only NGC~3576 diffuse mosaic shown) in red, H$\alpha$ from the MOSAIC~II camera (CTIO) in green, and {\em Spitzer}/IRAC 8~$\mu$m data in blue.  The green polygons represent extraction regions for diffuse X-ray emission. 
} 
\label{fig:cavity}
\end{center}
\end{figure}

NGC~3576 is another nearby, albeit less well-known, GHIIR, with D$\sim$2.8~kpc \citep{dePree99}.  It sits just $\sim 3.4^{\circ}$ east of the Carina star-forming complex and is part of the same Galactic spiral arm \citep{Georgelin00}.  Just $\sim 30\arcmin$ farther east lies NGC~3603 (Figures~\ref{fig:cavity}a and b), but this famous starburst cluster is thought to be much farther away, at D$\sim$7~kpc \citep{Rochau10}.  NGC~3603 is described in more detail below.  North of NGC~3576 is a region of low 8~$\mu$m surface brightness in the {\em MSX} data; we show a surface brightness contour in Figures~\ref{fig:cavity}a and b to emphasize this region and suggest that it might outline a cavity in the interstellar medium (ISM).

The massive stellar cluster ionizing the NGC~3576 GHIIR is still deeply embedded in its natal material and is located near the edge of a giant molecular cloud \citep[GMC,][]{Persi94}, shown in blue in Figure~\ref{fig:cavity}c.  It contains at least 51 stars earlier than A0, most with large IR excesses \citep{Maercker06}; not enough massive stars have been found to account for the strength of its radio emission \citep{Figueredo02, Barbosa03}.  A radio recombination line study of the ionized gas of the NGC~3576 nebula \citep{dePree99} revealed a north-south velocity gradient indicating a large-scale ionized outflow away from the embedded cluster.  This outflow may contribute to the prominent loops and filaments seen in H$\alpha$ (green in Figure~\ref{fig:cavity}c).  This new H$\alpha$ image of NGC~3576 was obtained with the MOSAIC~II camera \citep{Muller98} on the Cerro Tololo Inter-American Observatory (CTIO) Blanco 4m Telescope.  

\begin{figure}[htb] 
\begin{center}  
\includegraphics[width=1.0\textwidth]{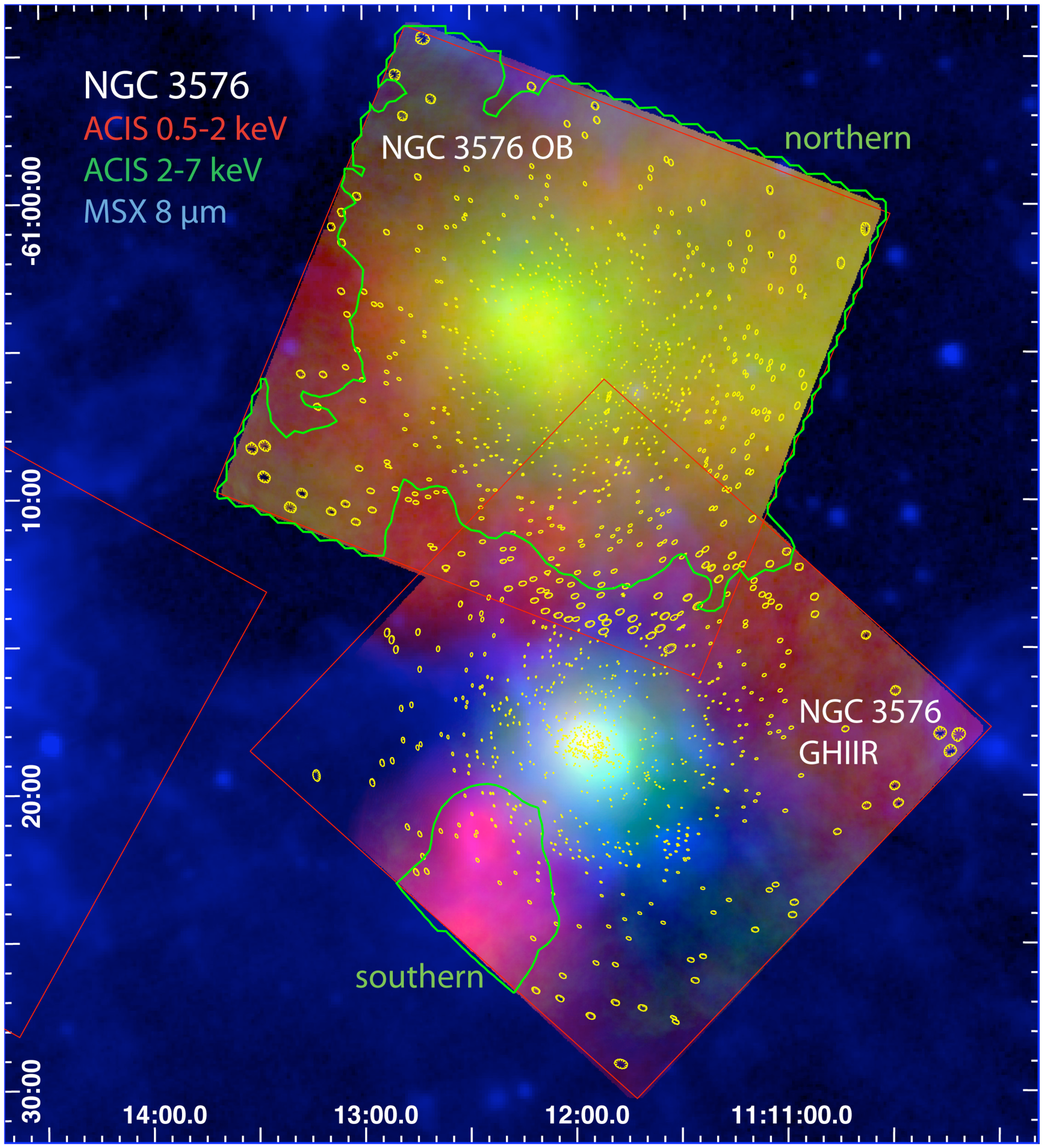}
\caption{Diffuse X-ray emission in NGC~3576.  For this target we show broad soft- and hard-band unresolved X-ray emission in the context of mid-IR emission from {\em MSX} (see Figure~\ref{fig:cavity}a).  ACIS point source extraction regions are shown in yellow and global diffuse extraction regions are shown in green.  The cavity north of NGC~3576 shows concentrated hard diffuse X-ray emission probably from a pulsar wind nebula, more extended hard diffuse emission perhaps from a recent cavity supernova, and ubiquitous soft diffuse emission that may fill the IR cavity outlined in Figures~\ref{fig:cavity}a and b above.  The ACIS-I pointing on NGC~3603 is indicated by a partial red box to the east of NGC~3576 (but the data are not shown).
} 
\label{fig:ngc3576image}
\end{center}
\end{figure}

Our ACIS-I observations consist of a 60-ks pointing on the embedded massive cluster \citep{Townsley09a} and another 60-ks pointing just to the north \citep{Townsley09b} (Figure~\ref{fig:cavity}, Figure~\ref{fig:ngc3576image}).  This northern pointing was designed to search for a young stellar cluster associated with the O8V+O8V eclipsing binary HD~97484 (EM~Car; placed at the ACIS-I aimpoint) and the O9.5Ib star HD~97319, the main members of the poorly-studied NGC~3576 OB Association \citep{Humphreys78}.  In this ACIS-I mosaic we have identified $>$1500 X-ray point sources, including a dense concentration of sources in the embedded cluster and several hundred more widely dispersed across the northern pointing and extending into the southern pointing (Figure~\ref{fig:cavity}b); their extraction regions are shown in yellow in Figure~\ref{fig:ngc3576image}.  This northern, comparatively loose concentration of X-ray sources around HD~97484 is the young stellar cluster we were hoping to find; its location in the IR cavity and its relaxed appearance compared to the young NGC~3576 embedded cluster lead us to speculate that it is several million years old.  HD~97484 and HD~97319 are likely its most massive remaining members.  The bright X-ray source just a couple of arcminutes northeast of HD~97484 is PSR~J1112-6103, a young ($<10^{5}$~yr), 65-ms pulsar \citep{Manchester01} of uncertain distance.  Its apparent proximity to this newly-discovered young cluster and its placement in the IR cavity suggest that it is at the distance of HD~97484 and NGC~3576 (2.8~kpc) and that it constitutes the remains of a more-massive O star that was also a member of this young cluster.  


The first ACIS-I observation of this target (the southern pointing) revealed a bright patch of soft diffuse X-ray emission near bright H$\alpha$ features southeast of the embedded cluster, along with extensive soft diffuse emission north of the edge of the GMC (Figure~\ref{fig:ngc3576image}).  The northern pointing yielded even more surprising diffuse X-ray emission:  a concentration of hard X-rays that may be the pulsar wind nebula of PSR~J1112-6103.  At a fainter level, hard X-rays nearly fill the field of view and extend into the southern pointing; they may be the signature of a cavity supernova remnant (possibly, although not necessarily, associated with PSR~J1112-6103).  The northern pointing also shows ubiquitous soft diffuse emission; the morphology of this soft emission indicates that it is either shadowed or displaced by the cold ISM that is traced by the {\em MSX} and {\em Spitzer} 8~$\mu$m emission.  There is no signature of the H$\alpha$ loops in the X-ray morphology.  The hot plasma sampled by these ACIS-I pointings very likely fills the IR cavity that we see in Figure~\ref{fig:cavity}.

\clearpage

\subsection{NGC~3603}

\begin{figure}[htb] 
\begin{center}  
\includegraphics[width=1.0\textwidth]{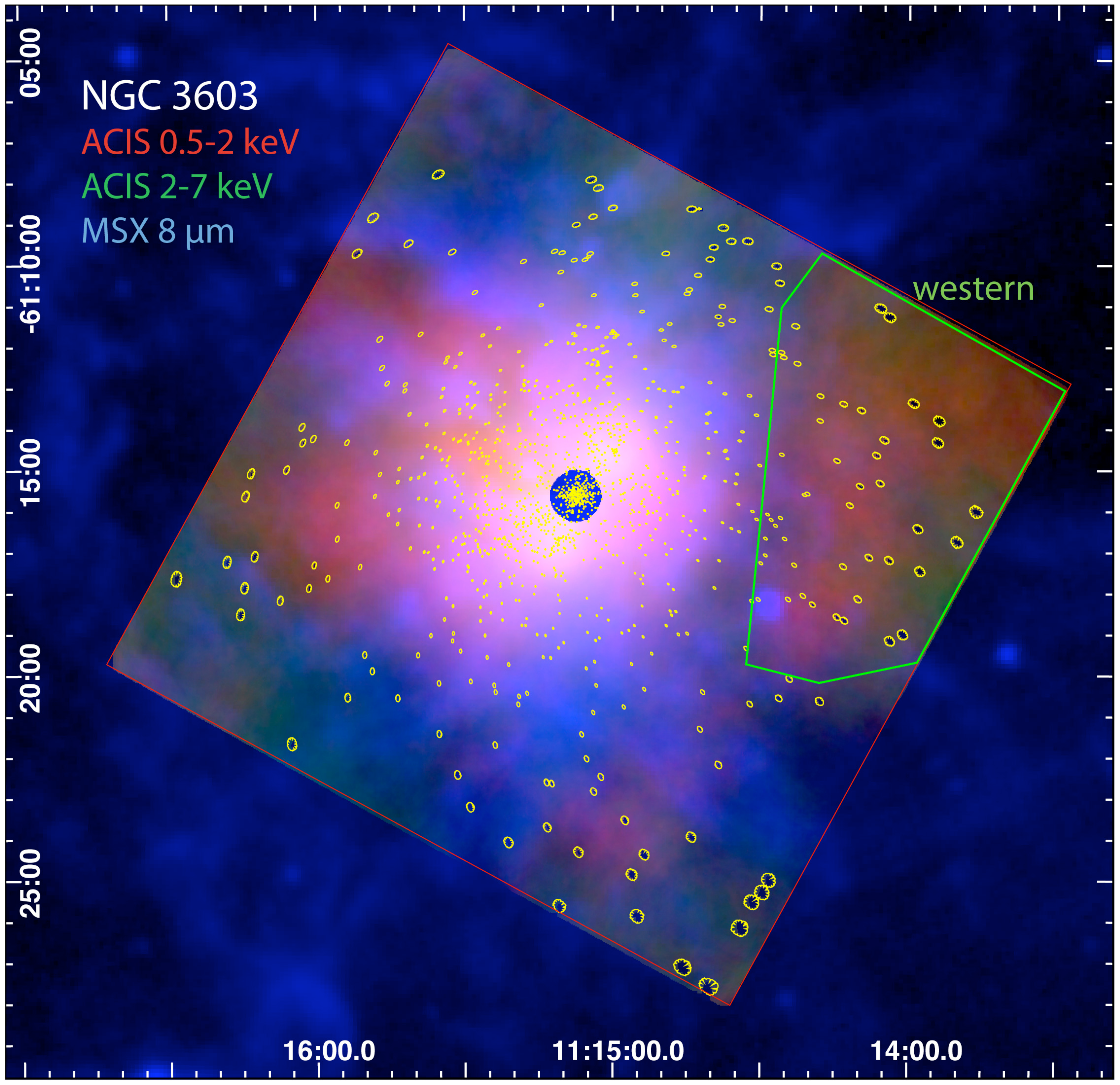}
\caption{Diffuse X-ray emission in NGC~3603.  Again we show broad soft- and hard-band unresolved X-ray emission in the context of mid-IR emission from {\em MSX} (see Figure~\ref{fig:cavity}a).  ACIS point sources are indicated by yellow polygons.  The X-ray emission was divided into two sections for spectral fitting:  the ``western'' region shown here and marked in Figure~\ref{fig:cavity}c, kept separate because it may be a foreground component, and the rest of the ACIS-I field.  The central blue circle indicates a region that was masked in the X-ray data for smoothing and spectral analysis because it is dominated by point sources at the center of the dense stellar cluster in NGC~3603.
} 
\label{fig:ngc3603image}
\end{center}
\end{figure}

NGC~3603's dense concentration of massive stars, ionization fronts, and dust pillars constitute the closest ``starburst cluster'' and it is often considered the Galactic analog of the ``super star cluster'' R136 in 30~Doradus.  It is the seventh most massive Galactic stellar cluster known \citep{Figer08}.  It is far less obscured ($A_{V}$$\sim$4.5~mag) than its more massive brethren \citep[e.g., Westerlund~1,][]{Clark08}, making NGC~3603 the most massive Galactic cluster accessible by visual telescopes.  Recent work with WFC3 on {\em HST} shows that the NGC~3603 complex has been forming stars for at least 10~Myr \citep{Beccari10}; this and the presence of evolved massive stars implies that this complex may well have seen cavity supernova activity.  The youngest stellar population in NGC~3603 is very young, with an age $\sim$1~Myr \citep{Rochau10}.  NGC~3603 provides an important contrast to Carina's ``cluster of clusters''; NGC~3603 contains as many O and WR stars as Carina but they are concentrated into a single custer covering $\sim$80~pc$^{2}$ rather than spread over $>$2200~pc$^{2}$ in multiple smaller clusters.  

NGC~3603 was observed in {\em Chandra} Cycle 1 (PI M.\ Corcoran) with 46~ks of usable time; \citet{Moffat02} found 384 ACIS-I sources but did not catalog them.  We have re-analyzed this observation and find 1328 X-ray point sources in the $17^{\prime} \times 17^{\prime}$ ACIS-I field -- a remarkable number for such a short observation of a distant region and a testament to the extraordinary stellar content of starburst clusters.  Clearly, though, most of the pre-MS X-ray point source population remains confused and unresolved in this short observation.

We have excised the ACIS point sources to study the diffuse X-ray emission; its spatial anti-coincidence with bright mid-IR structures tracing the surrounding heated dust (Figure~\ref{fig:ngc3603image}) is a clear indication of hot plasma filling the cavities carved out by massive stellar winds.  At the western edge of the NGC~3603 ACIS-I field, though, a large foreground (D $\sim$2.8~kpc) dust cavity is filled with diffuse X-rays (Figure~\ref{fig:cavity}), likely blown by the young stellar cluster that we described above as being associated with the NGC~3576 OB Association.  Due to this large and complicated foreground star-forming complex, it is unclear whether the faint diffuse X-rays seen in the NGC~3603 pointing are associated with NGC~3603 or with NGC~3576.  We attempt to untangle the diffuse X-ray emission from NGC~3603 and NGC~3576 by spectral fitting below.


\subsection{30~Doradus}

30~Doradus in the Large Magellanic Cloud (D $\sim$50~kpc) is the most luminous massive star-forming region in the Local Group.  At its center lies the 1--2 Myr-old ``super star cluster'' R136 \citep{Massey98}.  Although most work in 30~Dor concentrates on R136, the 30~Dor complex is $\sim$250~pc in diameter and contains at least five superbubbles, each several tens of parsecs across and filled with hot plasma emitting diffuse X-rays \citep{Chu90,Wang91}.  R136 sits at the confluence of at least three of these superbubbles.  It must be just the latest massive cluster to form; perhaps it is so massive because the actions of multiple superbubbles have concentrated the gas and triggered its violent star formation.

30~Dor's 50--100~pc superbubbles are well-known bright X-ray sources, where multiple cavity supernovae from past OB stars produce copious soft X-rays \citep{Chu90}.  R136 is too young to be the source of the supernovae that brighten the superbubbles in X-rays; rather these X-ray structures likely trace a variety of shocks caused by $\sim$100~Myr of massive star formation and evolution.  Past generations of massive clusters originally formed wind-blown bubbles that have grown into the superbubbles we now see via energy injection by supernovae.  Although we expect many of these massive clusters to have now largely dispersed, there must still be a reservoir of massive stars distributed throughout the 30~Dor complex, providing the fuel for ongoing off-center supernovae that Chu and Mac~Low proposed shock the superbubble walls and generate the bright diffuse X-rays that we see.  

30~Dor was one of the first targets observed by {\em Chandra}/ACIS-I; the diffuse X-ray emission seen in that early 22-ks observation was described by \citet{Townsley06a}.  We use that same dataset to characterize the diffuse X-ray emission in 30~Dor here; a later 90-ks ACIS-I observation will be described in a future paper.  Our \Chandra observations find several tens of 30~Dor's massive stars, most concentrated in R136 \citep{Townsley06b}, but the true extent of the remaining massive stars across the complex and their vast underlying young stellar populations remain largely unresolved in X-rays.

\begin{figure}[htb] 
\begin{center}  
\includegraphics[width=1.0\textwidth]{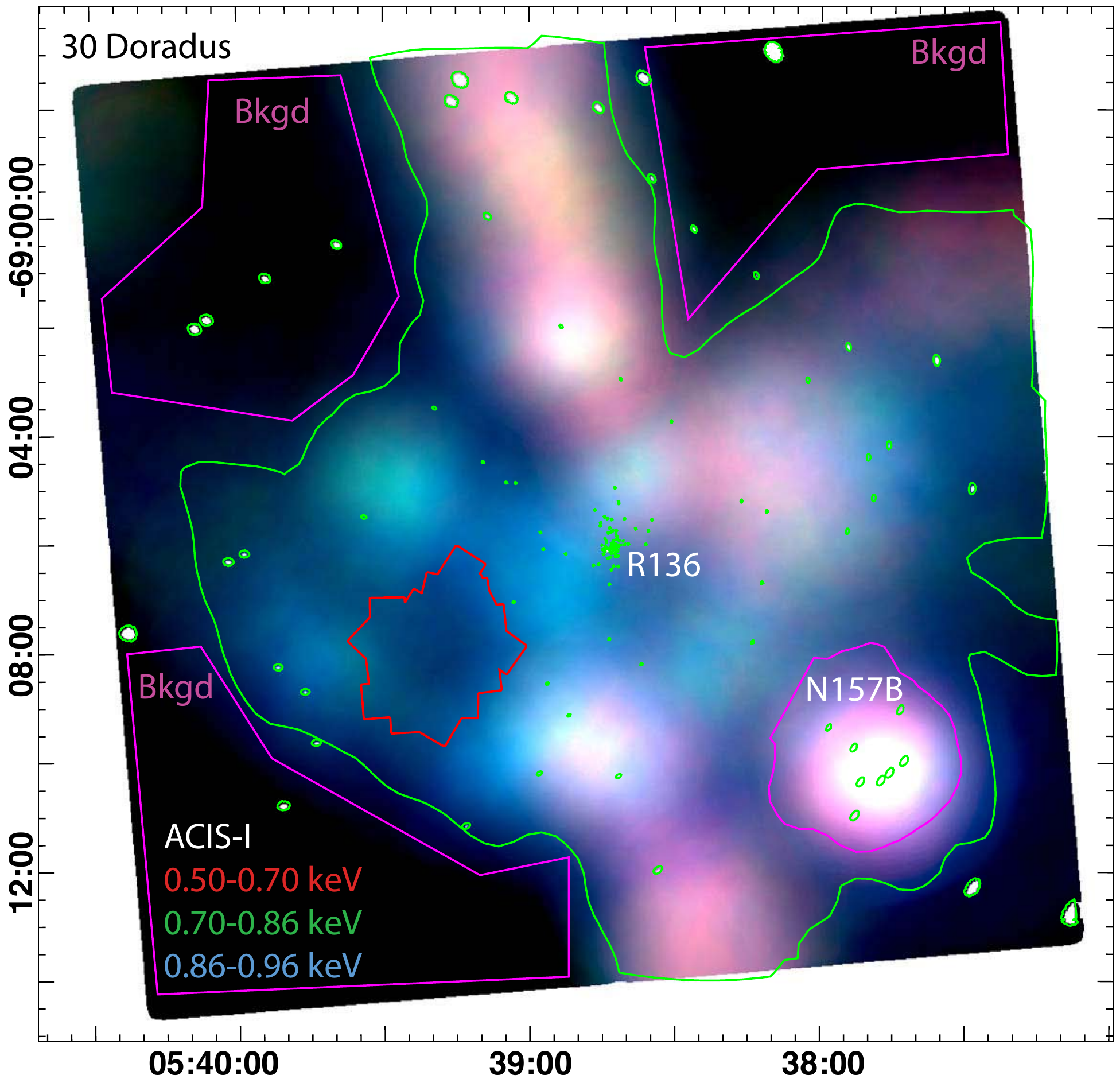}
\caption{Diffuse X-ray emission in 30~Doradus.  Here we show the three soft narrow-band ACIS-I images again, with red = 0.50--0.70~keV, green = 0.70--0.86~keV, and blue = 0.86--0.96~keV.  Large smoothing kernels are indicative of the small amount of data available here rather than a lack of underlying spatial complexity.  The region used to extract the diffuse spectrum is outlined in green.  For this target, background regions (shown in magenta) were used to create a background-subtracted diffuse spectrum.  As above, point sources have been masked before smoothing; their extraction regions are indicated in green.  The bright supernova remnant N157B and its pulsar (outlined in magenta) were also excluded from the diffuse emission spectrum.  For comparison, the red polygon shows the outline of the CCCP survey placed at the distance of 30~Dor.  The Carina complex would have roughly the same average surface brightness as the part of 30~Dor where we have placed the red polygon.
} 
\label{fig:30dorimage}
\end{center}
\end{figure}

Figure~\ref{fig:30dorimage} shows the same soft narrow-band smoothed ACIS images for 30~Dor that we showed for Carina and M17 in Figure~\ref{fig:carina+m17image}.  This short observation provides few photons in these narrow bands, so the adaptive-kernel smoothing scales are necessarily large.  This does not do justice to the complexity of the field; as more photon events are gathered in longer observations, more young stars will be resolved and more complexity in the diffuse emission will become apparent.  There are enough events here, however, to estimate the global diffuse X-ray properties of the complex.

\section{GLOBAL DIFFUSE X-RAY SPECTRA \label{sec:spectra}}

Using the global extraction regions for each target defined in the images above, we extracted integrated spectra for our massive star-forming regions, subtracted their instrumental backgrounds, then fit them in {\em XSPEC} \citep{Arnaud96} by starting with the complicated spectral model developed in the CCCP diffuse emission paper \citep{Townsley11b}.  The instrumental background was estimated using ACIS stowed data \citep{Hickox06}; this instrumental spectrum is subtracted from the global spectra before diffuse analysis begins \citep{Broos10}.

The CCCP diffuse spectral model consists of 3 non-equilibrium ionization (NEI) thermal plasma components ({\em vpshock} models in {\em XSPEC}, called kT1, kT2, and kT3 below) that describe the soft diffuse X-ray emission, plus harder thermal plasmas in collisional ionization equilibrium (CIE; {\em apec} models in {\em XSPEC}) that model unresolved point sources, both pre-main sequence (pre-MS) stars within the star-forming regions and background Galactic and extragalactic populations.  Details of the spectral extraction, background subtraction, and modeling process are described in \citet{Townsley11b}.  That paper explains that we chose NEI plasmas to model Carina's diffuse X-ray emission because we wanted to allow for the possibility of recent shocks contributing to the X-ray emission; such recent shocks might indicate recent cavity supernova activity or strong outflows from OB winds.  CIE conditions are recovered in the limit of long density-weighted timescales in NEI models; in Carina we found that the plasma components kT1 and kT3 often exhibited these long timescales and thus appear to be in CIE, but retaining the NEI model gives more flexibility to our spectral fitting and was found to be necessary for the kT2 component in Carina.  

Discussions of possible foreground diffuse emission components in the direction of Carina are also given in \citet{Townsley11b}; we expect such foreground components to contribute minimally to the integrated diffuse X-ray emission from our GHIIRs because the GHIIR X-ray emission often appears to be shadowed or displaced by colder ISM features commonly associated with star-forming regions and seen at longer wavelengths (e.g., pillars and ridges that glow in mid-IR PAH emission), but sometimes our models include minimally-obscured soft plasma components that could be due to foreground emission.  This issue will be explored in future, detailed studies of the individual GHIIRs.  Please note that ACIS-I has minimal response below 0.5~keV, thus our ability to constrain the value of low absorbing columns or the temperatures of very soft thermal plasmas is quite limited.

A complicated spectral model is necessary for the Galactic star-forming regions described here because no part of the ACIS-I field of view can be guaranteed to be free of diffuse X-ray emission, so there is no part of the field that can be called ``background'' and defined as such for spectral fitting.  In 30~Dor the situation is different; as described above, parts of the field show minimal diffuse flux and can be treated as samples of the background X-ray emission and used to subtract off the background components, leaving only the soft diffuse X-ray emission to model.

In \citet{Townsley11b}, we found that no combination of NEI and CIE plasmas could adequately fit the tessellated spectra describing Carina's diffuse X-ray emission.  Fit residuals showed prominent unmodeled lines, often close in energy to the emission lines expected from hot plasmas, but different enough that no combination of hot plasma models could provide fits to these lines.  They were well-fit by narrow gaussians, though, and sometimes similar in energy to unmodeled lines found by \citet{Ranalli08} in thermal plasma fits to {\em XMM-Newton} spectra of the starburst galaxy M82.  Ranalli et al.\ proposed that these lines come from charge exchange (CE) between M82's hot galactic superwind and dust in its cold neutral clouds.  \citet{Lallement04} summarizes the CE process:  when hydrogenic or helium-like ions from a hot plasma strike neutral clouds, electrons freed from the neutrals are captured by the ions into high-excitation states; subsequent continuum-free line emission from both the ions and neutrals (which presumably have recaptured another electron into a high-excitation state) results.  Lallement explains that the CE process is important for low-density hot plasmas hitting high-density cold clouds, e.g., for galactic superwinds hitting cold halo clouds, thus we surmised that CE also might be a plausible explanation for the unmodeled narrow emission lines seen in Carina's diffuse X-ray spectra.  

We note that we know of no previous claim (prior to our CCCP study) for the detection of X-ray CE emission lines in \hii regions and that CE may not in fact be the X-ray emission mechanism responsible for the unmodeled spectral features that we saw in Carina and that we see below in other GHIIRs.  With this caveat in mind and for the sake of brevity in the descriptions and discussion to follow, though, we will refer to this phenomenon as charge exchange and hope that these observations spur the community to investigate the true nature of this emission.

\begin{figure}[htb] 
\begin{center}  
\includegraphics[width=0.4\textwidth]{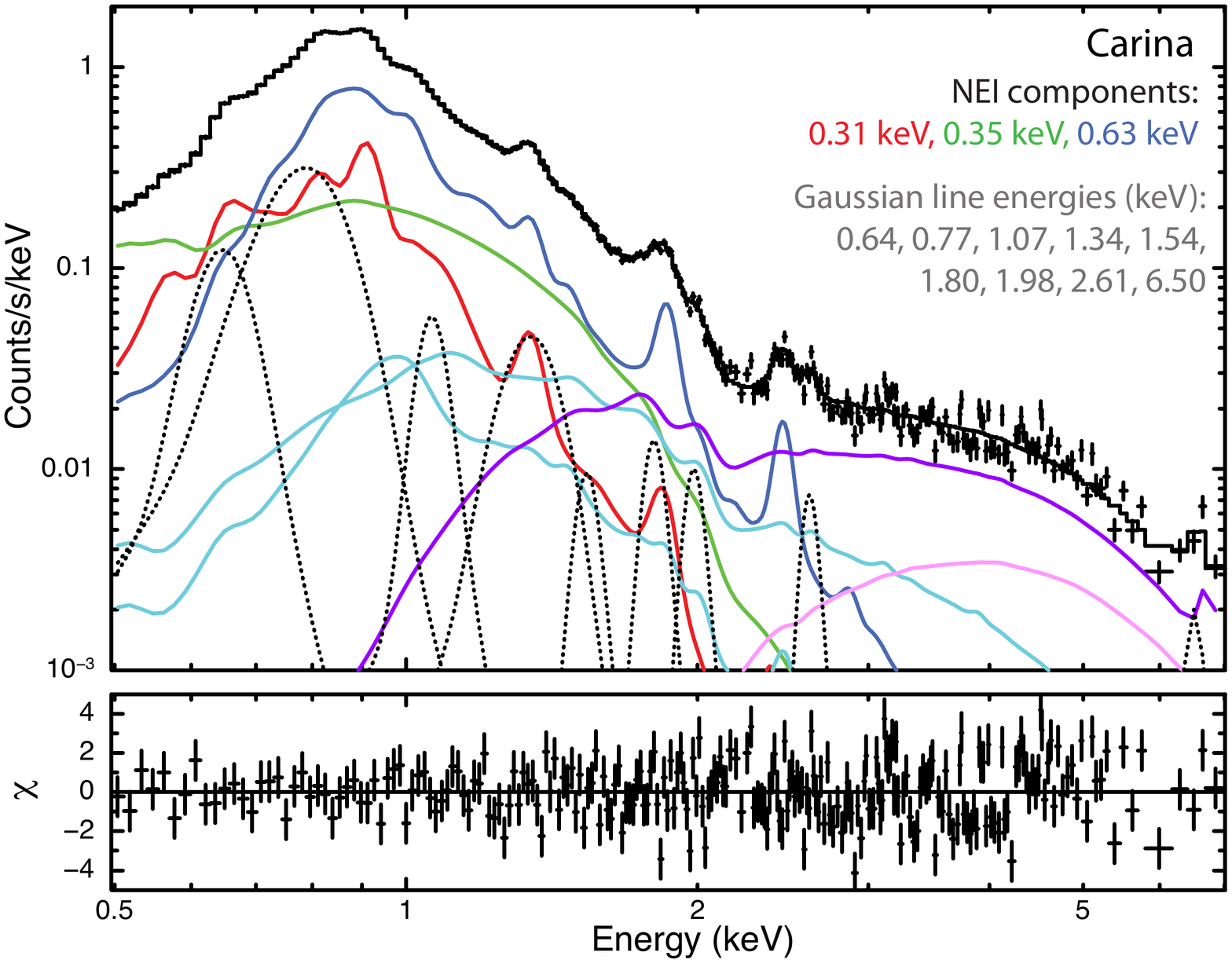}
\includegraphics[width=0.4\textwidth]{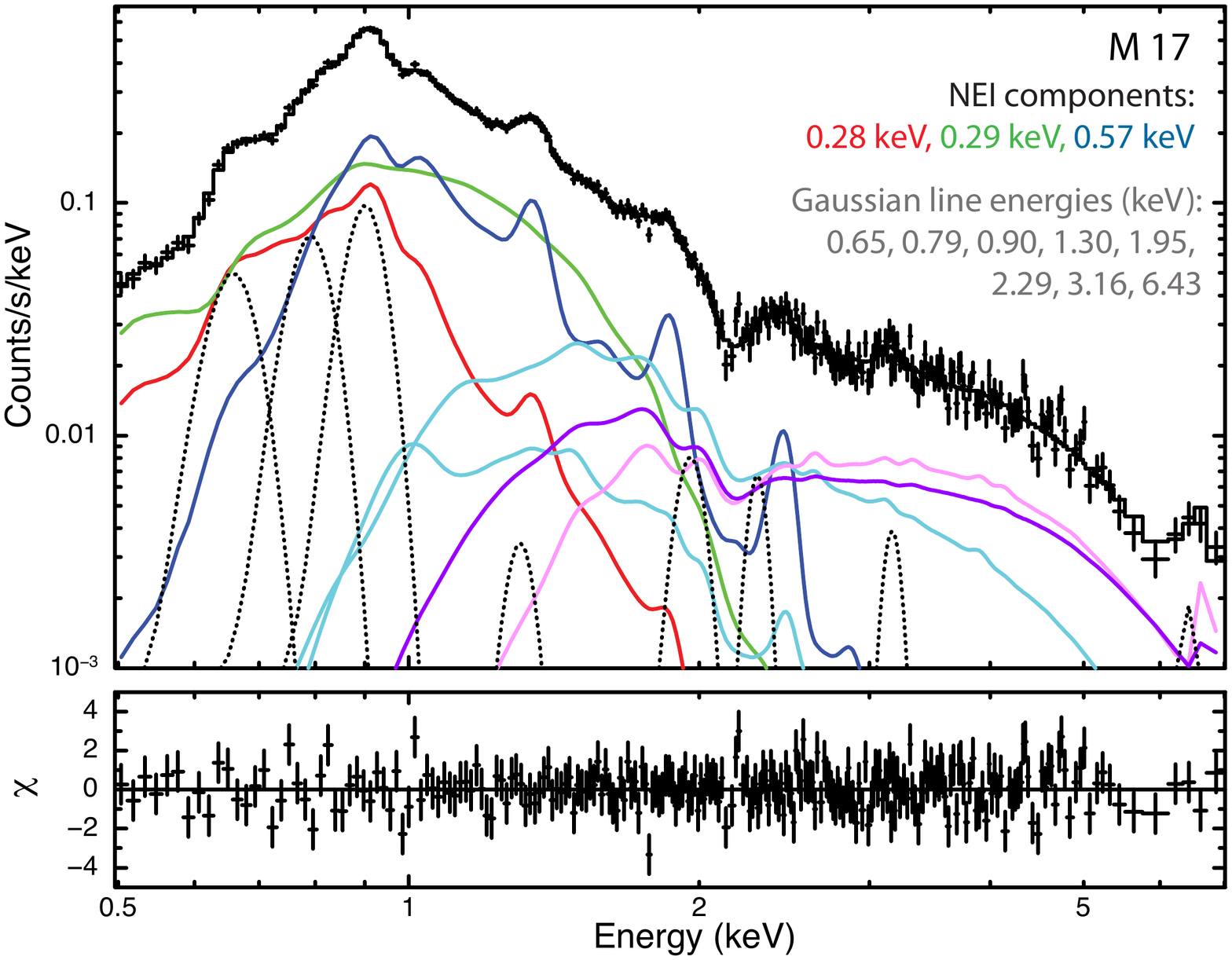}
\includegraphics[width=0.4\textwidth]{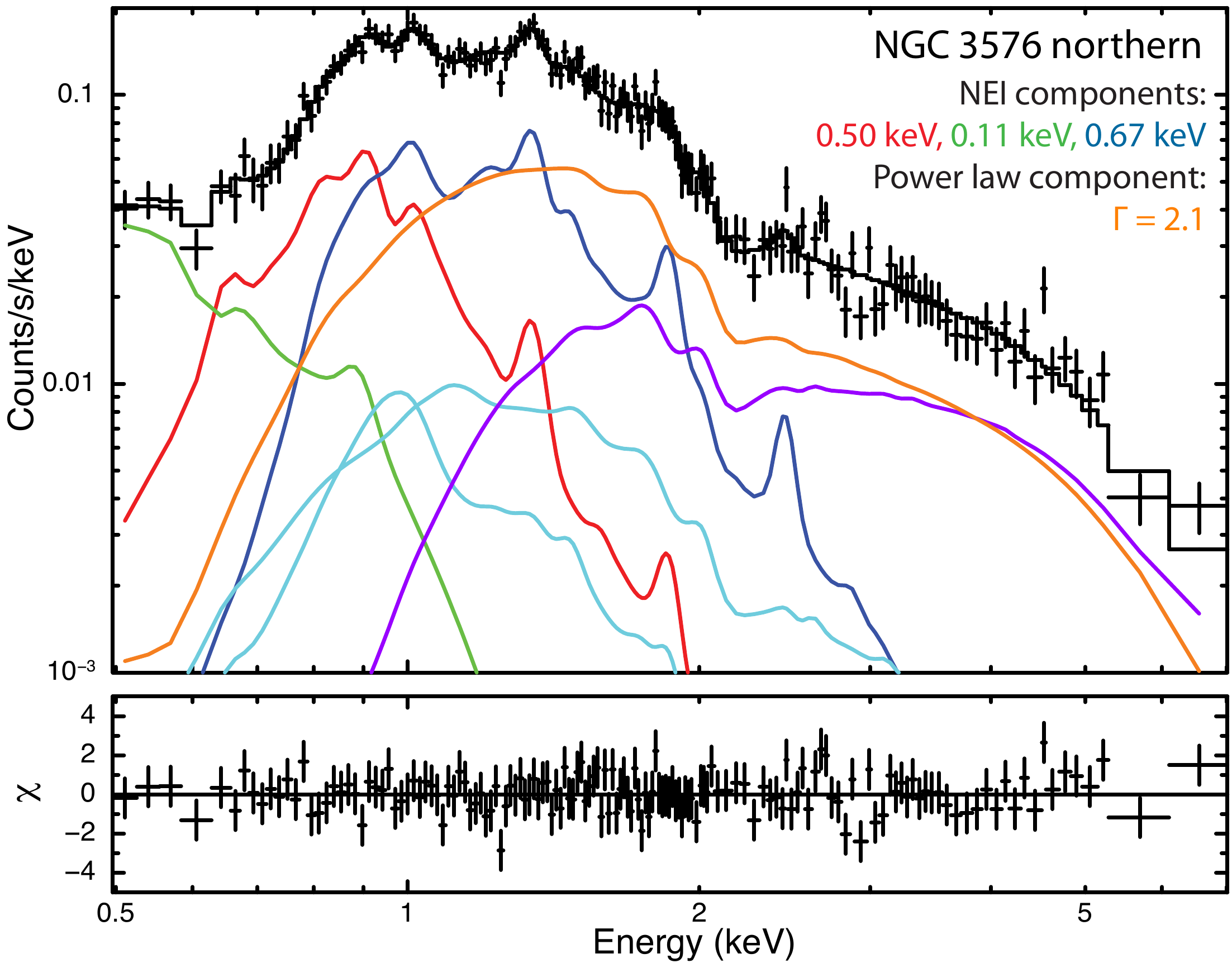}
\includegraphics[width=0.4\textwidth]{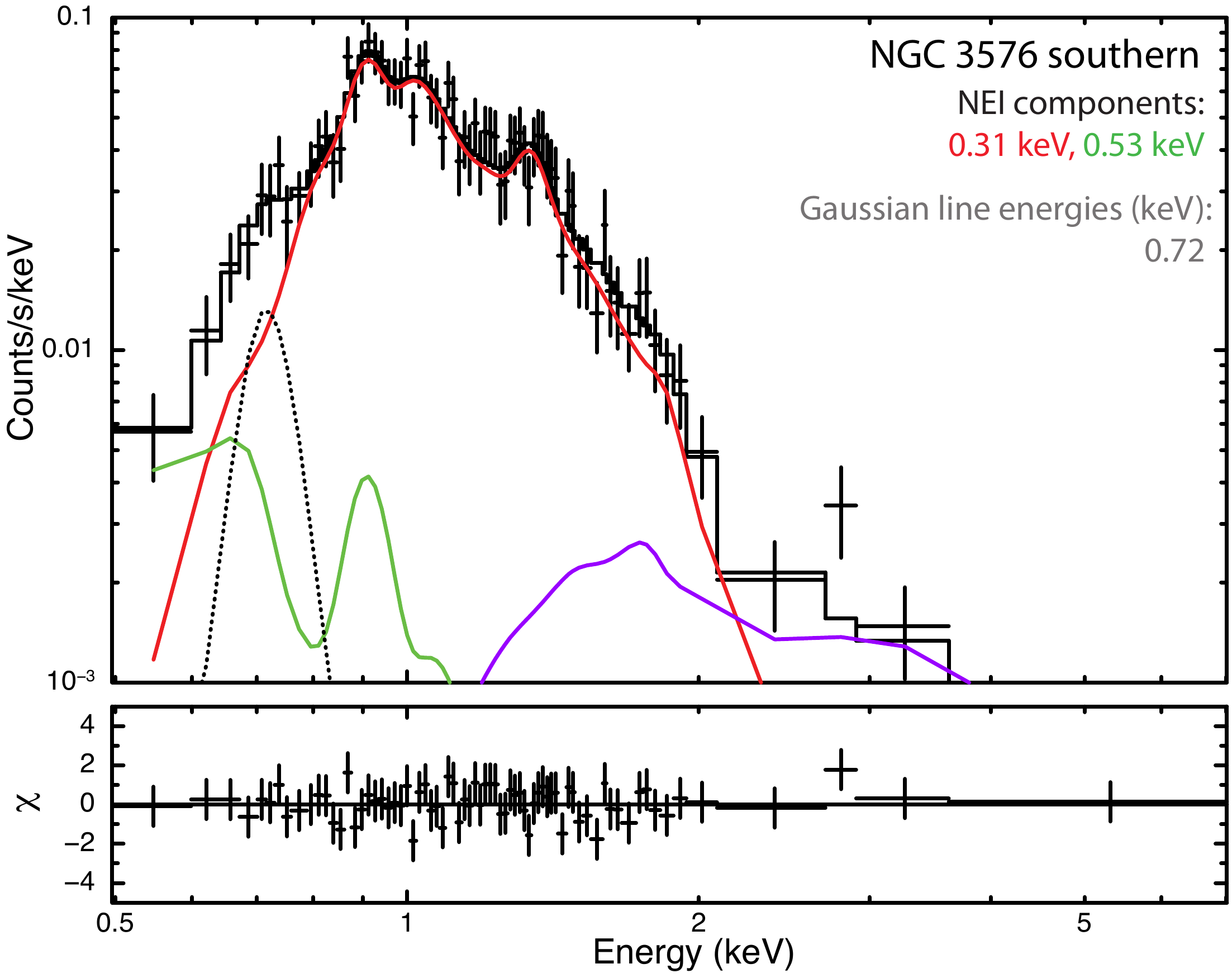}
\includegraphics[width=0.4\textwidth]{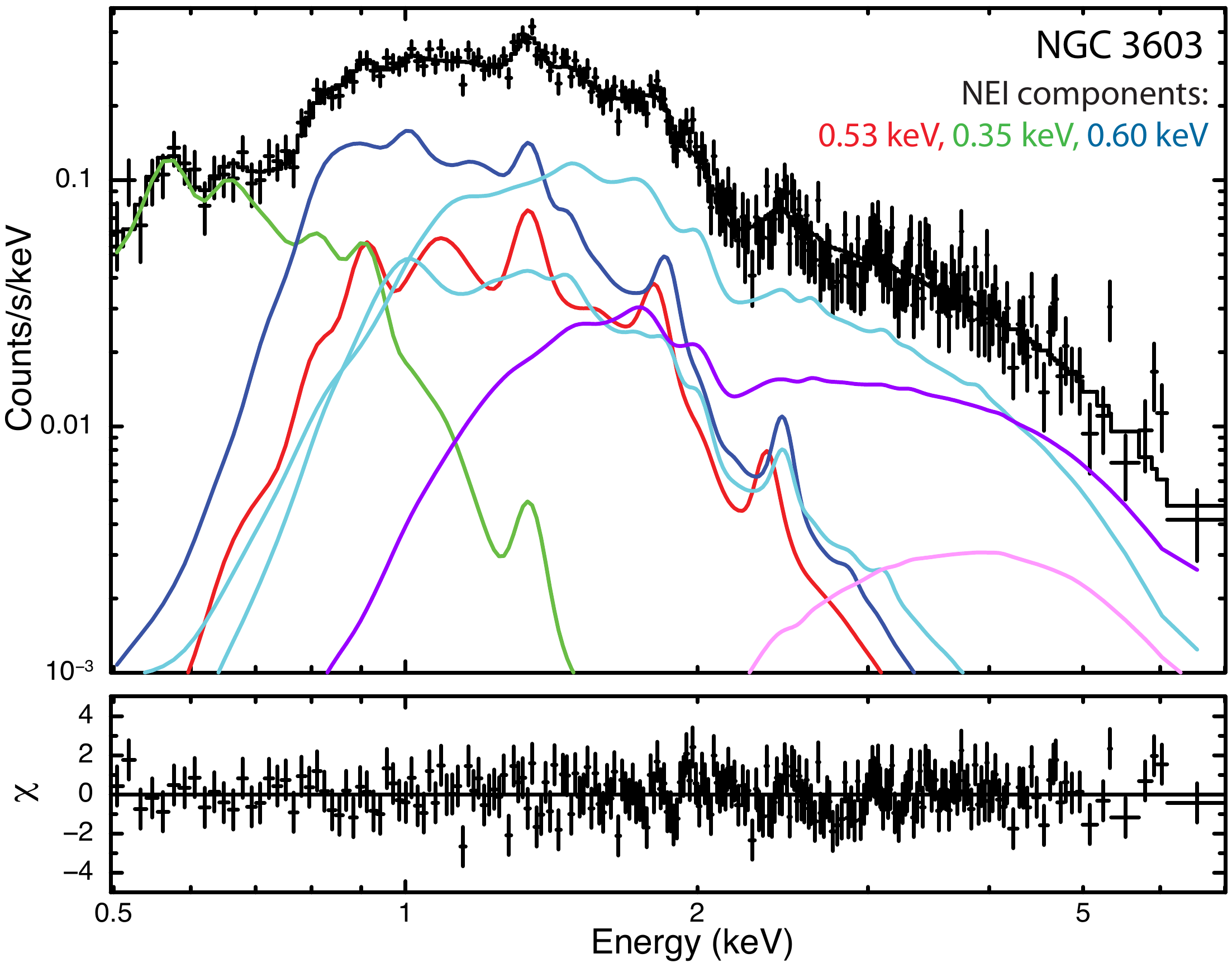}
\includegraphics[width=0.4\textwidth]{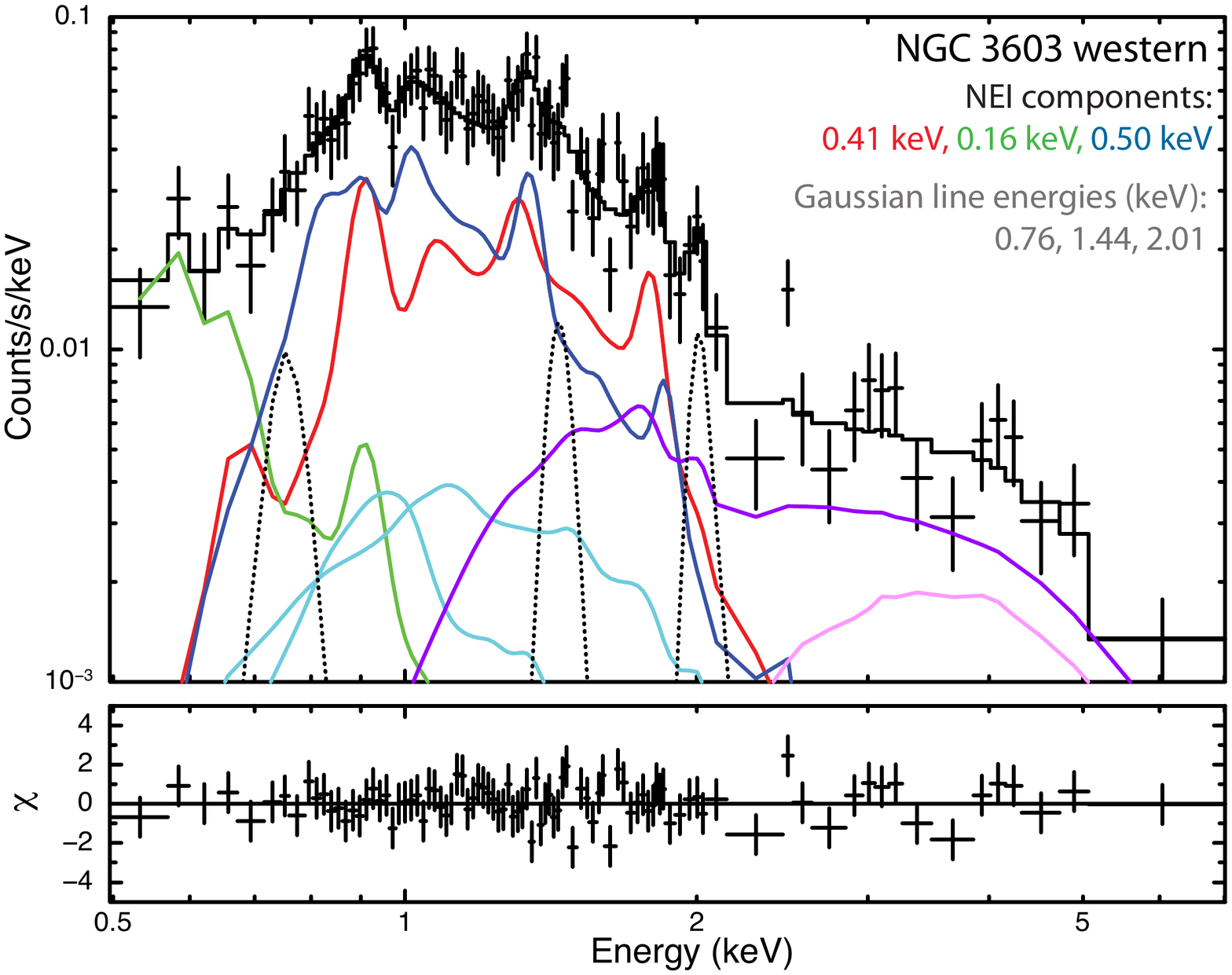}
\includegraphics[width=0.4\textwidth]{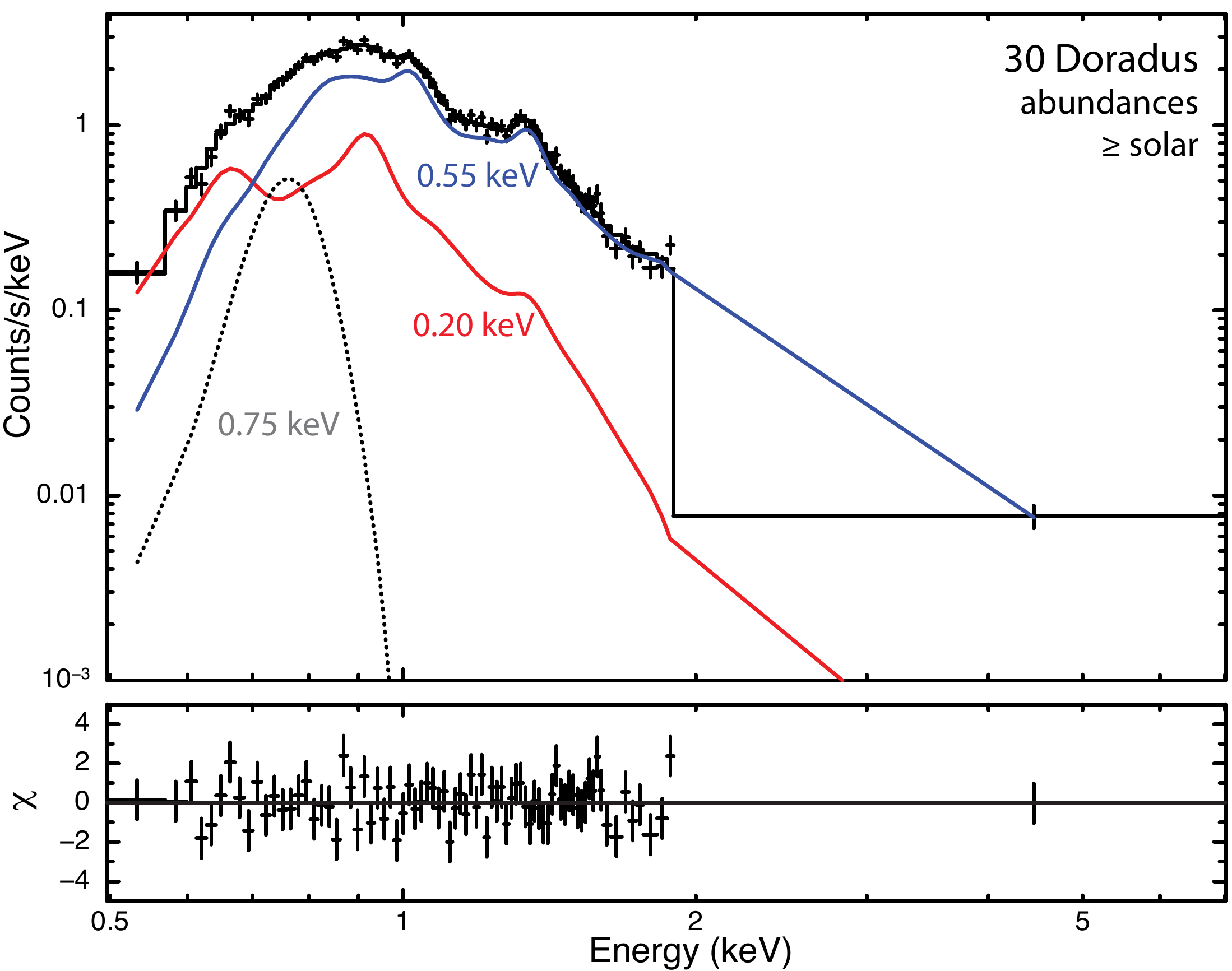}
\includegraphics[width=0.4\textwidth]{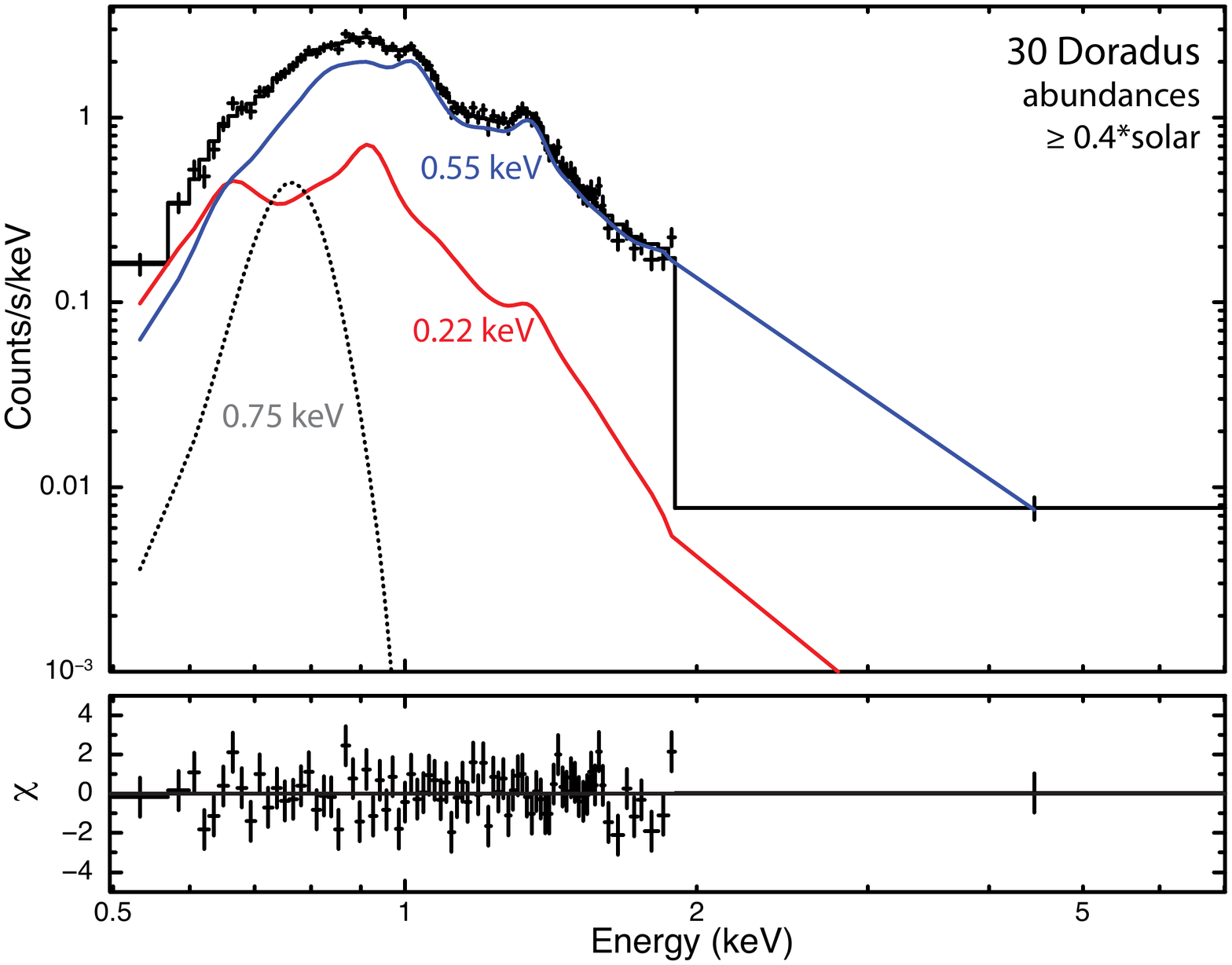}
\caption{Global diffuse X-ray spectra; please note that ordinate axis ranges vary between spectra and zooming reveals more detail.  The 3 NEI thermal plasma components presumably tracing diffuse emission ({\em vpshock} models called kT1, kT2, and kT3 in the text and in Table~\ref{tbl:diffuse_spectroscopy}) are shown in red, green, and dark blue, respectively.  Gaussians that improved the fit are shown as dotted grey lines.  ``Nuisance'' components due mainly to unresolved pre-MS stars (cyan), Galactic Ridge emission (purple), and unresolved extragalactic sources (pink) are shown where necessary.  Other details are described in the text.
} 
\label{fig:spectra}
\end{center}
\end{figure}

The integrated diffuse spectra for all GHIIRs studied here, and their model fits, are shown in Figure~\ref{fig:spectra}; see the annotations in each panel for the most relevant diffuse component fit parameters.  We show all of these complicated models in a single figure to facilitate comparison between them; each spectrum is described in turn below and in Table~\ref{tbl:diffuse_spectroscopy}, which gives the diffuse emission spectral fit parameters and important derived quantities.

\clearpage

The cyan curves in Figure~\ref{fig:spectra} represent two thermal plasma components that model unresolved pre-MS star emission; in the COUP study, pre-MS stars were found to have X-ray spectra that were well-described by such a model, with a soft component at $\sim$0.86~keV and a hard component at $\sim$2.6~keV \citep{Preibisch05}.  The {\em XSPEC} \citep{Arnaud96} model form used for this constituent is {\em TBabs(apec + apec)}, with abundances set at 0.3Z$_{\odot}$ to approximate pre-MS star abundances and the normalization of the soft component constrained to be half that of the hard component, as found by Preibisch et al.\ for the COUP sample (the normalization for the hard component is free to vary).  In \citet{Townsley11b} we represented the hard component of the unresolved stellar emission by a single absorbed thermal plasma ({\em TBabs*apec}) and did not attempt to separate out the unresolved pre-MS stars' soft component from diffuse emission components in the model, but accounted for it after the fact by modeling the faint population of resolved point sources.  This change in the model form more clearly separates unresolved stellar emission from diffuse emission because we can require both pre-MS star emission components to have the same absorbing column.  

The purple and pink curves in Figure~\ref{fig:spectra} represent the same background components (Galactic Ridge emission and unresolved extragalactic sources, primarily AGN) and use the same spectral model components ({\em TBabs*apec}) as the Carina diffuse emission model described in \citet{Townsley11b}.  The Galactic Ridge component (purple) was fixed in absorption ($N_{H} = 2.0 \times 10^{22}$~cm$^{-2}$), temperature ($kT = 10$~keV), and surface emission measure (emission measure per unit area, $SEM = 8.07 \times 10^{53}$~cm$^{-3}$~pc$^{-2}$) to the Carina value for all Galactic star-forming regions.  The model component representing unresolved extragalactic sources (pink) is allowed to vary, so it could also be taking up any variations in Galactic Ridge emission in these other fields.  When it is required at all, it shows high absorption ($N_{H} > 4 \times 10^{22}$~cm$^{-2}$) and a hard thermal plasma ($kT > 4$~keV).

\setlength{\tabcolsep}{0.03in}
\renewcommand{\arraystretch}{2.0}
\begin{deluxetable}{@{}lrrr*{3}{l}*{3}{l}*{3}{r}*{3}{c}*{2}{l}r*{4}{c}@{}}   

\centering \rotate \tabletypesize{\tiny} \tablewidth{0pt}

\tablecaption{Spectral Fits for Global Diffuse Emission \label{tbl:diffuse_spectroscopy}}

\tablehead{
\multicolumn{4}{c}{Target\tablenotemark{a}} &
\multicolumn{15}{c}{Selected Spectral Fit Parameters\tablenotemark{b}} &
\multicolumn{4}{c}{Inferred Diffuse Emission\tablenotemark{c}} \\ 
\multicolumn{4}{c}{\hrulefill} &
\multicolumn{15}{c}{\hrulefill} &
\multicolumn{4}{c}{\hrulefill}\\
\multicolumn{4}{c}{Diffuse Region} &
\multicolumn{3}{c}{NEI Absorptions} &     
\multicolumn{3}{c}{NEI Temperatures} &     
\multicolumn{3}{c}{NEI Timescales} &     
\multicolumn{3}{c}{NEI ``Surface'' EMs} &     
\colhead{$>Z_{\odot}$} & 
\colhead{Gau, $\Gamma$} &
\colhead{Quality} &
\multicolumn{4}{c}{Surface Brightnesses}  \\
\multicolumn{4}{c}{\hrulefill} &
\multicolumn{3}{c}{\hrulefill} &
\multicolumn{3}{c}{\hrulefill} &
\multicolumn{3}{c}{\hrulefill} &
\multicolumn{3}{c}{\hrulefill} &
\multicolumn{1}{c}{\hrulefill} &
\multicolumn{1}{c}{\hrulefill} &
\multicolumn{1}{c}{\hrulefill} &
\multicolumn{4}{c}{\hrulefill}   \\
\colhead{Name} & \colhead{D} & \colhead{Area} & \colhead{NetCts} &
\colhead{$N_{H1}$} & \colhead{$N_{H2}$} & \colhead{$N_{H3}$} & 
\colhead{kT1}      & \colhead{kT2}      & \colhead{kT3}      & 
\colhead{${\tau}1$}& \colhead{${\tau}2$}& \colhead{${\tau}3$}&  
\colhead{SEM1}     & \colhead{SEM2}     & \colhead{SEM3}     &  
\colhead{} &  \colhead{} &     
\colhead{$\chi^2 \!/\! \rm{DOF}$} &
\colhead{SB1$_{tc}$} & \colhead{SB2$_{tc}$} & \colhead{SB3$_{tc}$} & \colhead{SB$_{\rm{gau,\Gamma}}$} 
\\
\colhead{} & \colhead{(kpc)} & \colhead{(pc$^2$)} & \colhead{} &
\multicolumn{3}{c}{($\log$ cm$^{-2}$)} & \multicolumn{3}{c}{(keV)} & \multicolumn{3}{c}{($\log$ cm$^{-3}$ s)} & \multicolumn{3}{c}{($\log$ cm$^{-3}$ pc$^{-2}$)} & 
&  
\colhead{(keV)} &
&
\multicolumn{4}{c}{($\log$ erg s$^{-1}$ pc$^{-2}$)} 
\\
\numberthecolumn & \numberthecolumn & \numberthecolumn & \numberthecolumn &
\numberthecolumn & \numberthecolumn & \numberthecolumn & 
\numberthecolumn & \numberthecolumn & \numberthecolumn & 
\numberthecolumn & \numberthecolumn & \numberthecolumn & 
\numberthecolumn & \numberthecolumn & \numberthecolumn & \numberthecolumn & \numberthecolumn &
\numberthecolumn &
\numberthecolumn & \numberthecolumn & \numberthecolumn & \numberthecolumn 
\setcounter{column_number}{1}
}



\startdata

Carina  &  2.3 & 1207 & 1.012e6 &
$21.6\phd_{-0.04}^{+0.14}$ & $21.0\phd_{-0.02}^{+0.39}$ & $21.1\phd_{-0.03}^{+0.09}$  & 
$0.31\phd$                 & $0.35\phd_{-0.00}^{+0.03}$ & $0.63\phd_{-0.03}^{+0.01}$  &
$11.2\phd$  & $8.6\phd$  & $12.9\phd$  & 
$54.4\phd_{-0.1}^{+0.1}$  & $55.2\phd_{-0.1}^{+0.1}$ & $54.1\phd_{-0.1}^{+0.1}$    & 
O, Ne, & 
0.64, 0.77, 1.07, & 
548/214 &  
31.91 &   31.34 &   31.46 &   30.98 \\
  &  &  &  &  &  &  &  &  &  &  &  &  &  & & & Si, S, & 1.34, 1.54, 1.80, &  &  &  &  & \\ 
  &  &  &  &  &  &  &  &  &  &  &  &  &  & & & Fe     & 1.98, 2.61, 6.50  &  &  &  &  & \\

M17  &  2.1 & 85.6 & 152,000 &
$21.2\phd_{\cdots}^{+0.7}$ & $21.7\phd_{-0.04}^{+0.19}$ & $22.0\phd_{-0.03}^{\cdots}$  & 
$0.28\phd_{-0.09}^{+0.09}$ & $0.29\phd_{-0.04}^{+0.08}$ & $0.57\phd_{-0.04}^{+0.15}$  &
$13.7\phd$  & $8.8\phd$  & $11.1\phd$  & 
$54.0\phd_{-0.5}^{+0.3}$  & $55.8\phd_{-0.2}^{+0.1}$ & $54.5\phd_{-0.1}^{+0.4}$  & 
Ne, Si, & 
0.65, 0.79, 0.90, & 
263/218 &  
31.07 &   31.91 &   32.12  &   30.88 \\
  &  &  &  &  &  &  &  &  &  &  &  &  &  & & & S & 1.30, 1.95, 2.29, &  &  &  &  &  \\
  &  &  &  &  &  &  &  &  &  &  &  &  &  & & &   & 3.16, 6.43  &  &  &  &  & \\

NGC~3576~S & 2.8 & 15.6 & 2600 &
$22.1\phd_{-0.09}^{+0.06}$ & $21.4\phd$   & $\cdots\phd$  & 
$0.31\phd_{-0.07}^{+0.06}$ & $0.53\phd$   & $\cdots\phd$  &
$12.2\phd$  & $9.7\phd$  & $\cdots\phd$  & 
$55.4\phd_{-0.1}^{+0.2}$  & $53.3\phd_{-1.3}^{+1.2}$ & $\cdots\phd$  & 
O, Ne & 
0.72  & 
46/59 &  
32.76 &   31.12 &   $\cdots\phd$ & 32.06 \\

NGC~3576~N &  2.8 & 173.8 & 25,000 &
$21.5\phd$                 & $21.5\phd$                 & $22.1\phd_{-0.2}^{\cdots}$  & 
$0.50\phd_{-0.4}^{\cdots}$ & $0.11\phd_{\cdots}^{+0.1}$ & $0.67\phd_{-0.1}^{\cdots}$ &
$11.4\phd$  & $9.3\phd$  & $12.3\phd$  & 
$53.4\phd_{-0.4}^{+0.1}$ & $56.1\phd_{-1.3}^{+0.4}$ & $54.3\phd_{-0.3}^{+0.2}$    & 
$\cdots\phd$  & 
$\Gamma = 2.12_{\cdots}^{+0.36}$  & 
131/136 &  
30.86 &   31.08 &   31.53 &   31.23 \\

NGC~3603 & 7.0 & 956.0 & 22,700 &
$22.3\phd_{-1.0}^{\cdots}$ & $20.2\phd_{\cdots}^{+0.9}$ & $22.1\phd_{-0.11}^{+0.08}$  & 
$0.53\phd_{-0.3}^{\cdots}$ & $0.35\phd$                 & $0.60\phd_{-0.07}^{+0.33}$   &
$10.3\phd$  & $10.9\phd$  & $13.7\phd$  & 
$54.7\phd_{-0.4}^{+0.3}$ & $53.1\phd_{-0.2}^{+0.4}$ & $54.4\phd_{-0.1}^{+0.2}$    & 
$\cdots\phd$  & 
$\cdots\phd$  & 
226/213 &  
32.37 &   30.52 &   31.54 & $\cdots\phd$ \\

NGC~3603~W & 7.0 & 212.2 & 3600 &
$22.2\phd_{-0.1}^{\cdots}$ & $20.8\phd$                 & $22.1\phd_{-0.08}^{+0.10}$  & 
$0.41\phd_{-0.2}^{+0.1}$   & $0.16\phd$                 & $0.50\phd_{\cdots}^{+0.5}$   &
$9.8\phd$  & $13.7\phd$  & $11.7\phd$  & 
$55.3\phd_{-0.5}^{+0.2}$ & $53.8\phd_{-0.5}^{+0.4}$ & $54.4\phd_{-0.3}^{+0.2}$    & 
$\cdots\phd$  & 
0.76, 1.44, 2.01 & 
75/71 &  
32.67 &   30.49 &   31.77  & 31.29 \\

30~Dor solar+ & 50.0 & 33389 & 32,400 &
$21.84\phd_{-0.02}^{+0.05}$   & $\cdots\phd$   & =$N_{H1}\phd$  & 
$0.20\phd_{-0.02}^{+0.04}$ & $\cdots\phd$             & $0.55\phd_{-0.04}^{+0.03}$      &
$13.7\phd$  & $\cdots\phd$  & $13.0\phd$  & 
$ 56.3\phd_{-0.5}^{+0.1}$ & $\cdots\phd$ & $ 56.1\phd_{-0.1}^{+0.1}$    & 
O, Ne, & 
0.75  &
86/66 &  
32.50 &   $\cdots\phd$ &   32.32  & 31.54 \\
 &  &  &  &  &  &  &  &  &  &  &  &  &  & & & Mg, Fe &  &  &  &  &  & \\

30~Dor subsolar & 50.0 & 33389 & 32,400 &
$21.77\phd_{-0.10}^{+0.09}$   & $\cdots\phd$   & =$N_{H1}\phd$  & 
$0.22\phd_{-0.04}^{+0.13}$ & $\cdots\phd$             & $0.55\phd_{-0.06}^{+0.06}$      &
$13.7\phd$  & $\cdots\phd$  & $12.3\phd$  & 
$ 56.6\phd_{-0.6}^{+0.6}$ & $\cdots\phd$ & $ 56.5\phd_{-0.3}^{+0.1}$    & 
O, Ne, & 
0.75  & 
89/67 &  
32.17 &   $\cdots\phd$ &   32.30  & 31.33 \\
 &  &  &  &  &  &  &  &  &  &  &  &  & & & & Mg &  &  &  &  &  & \\
\enddata

               
\tablenotetext{a}{
Target details.  
\\Col.\ (1):  Target name.
\\Col.\ (2):  Assumed distance. 
\\Col.\ (3):  Geometric area of the diffuse extraction region, irrespective of point source masking. 
\\Col.\ (4):  Net counts in global diffuse spectrum.
}

\tablenotetext{b}{
All {\em XSPEC} fits used the {\em source} model ``TBabs*vpshock + TBabs*vpshock + TBabs*vpshock + TBabs(apec + apec) + TBabs*apec + TBabs*apec,'' similar to that developed for Carina's diffuse emission \citep{Townsley11b} but modified slightly to include a 2-component model for unresolved pre-MS stars, with gaussian lines or power law components added as necessary (described in the text).  Large numbers are reported logarithmically.
\\Col.\ (5)-(7): The best-fit value for the extinction column density ({\it TBabs} components).
\\Col.\ (8)-(10): The best-fit value for the plasma temperature ({\it vpshock} components).
\\Col.\ (11)-(13): The density-weighted ionization timescale for the plasma ({\it vpshock} components).
\\Col.\ (14)-(16): The ``surface'' emission measure (emission measure per unit area) for the plasma ({\it vpshock} components).  This quantity is invented here to allow comparison between regions of different area.
Uncertainties represent 90\% confidence intervals.
More significant digits are used for uncertainties $<$0.1 in order to avoid large rounding errors; for consistency, the same number of significant digits is used for both lower and upper uncertainties.
Uncertainties are missing when XSPEC was unable to compute them or when their values were so large that the parameter is effectively unconstrained.  
NEI density-weighted timescales are reported without errors because {\it XSPEC} could rarely determine these errors; these timescales should be treated as order-of-magnitude estimates only.
\\Col.\ (17): Supersolar abundances.  Abundances were tied together in the 3 NEI models and were frozen at 1.0 $Z_\odot$ unless otherwise indicated.  Abundances of O, Ne, Mg, Si, S, and Fe have values greater than solar in some fits; this is reported here but the actual abundance values are omitted because they are not well-constrained.  The Ni abundance was tied to Fe.
\\Col.\ (18): Other important diffuse model components.  Line energies of added gaussians are shown; 90\% confidence intervals on these values are typically less than $\pm 20$~eV, but they can approach $\pm 100$~eV for the highest-energy lines.  The power law slope is given for the NGC~3576 northern region.
}

\tablenotetext{c}{
All inferred intrinsic diffuse plasma properties assume the distances in Col.\ (2) and are given for the total band, 0.5--7~keV.
\\Cols.\ (20)-(22): Absorption-corrected total-band surface brightness (intrinsic luminosity per unit area) of the diffuse emission ({\it vpshock} components).
\\Col.\ (23):  Absorption-corrected total-band surface brightness for all gaussian lines combined, except for NGC~3576 north, where this is the absorption-corrected surface brightness for the power law component.
}

\end{deluxetable}  

\subsection{Carina}

The integrated diffuse X-ray emission spectrum of Carina has over a million counts in the ``total'' band (0.5--7~keV) and covers $\sim$2697 square arcminutes.  It is distinctive from other global spectra shown in Figure~\ref{fig:spectra} in that it clearly shows a strong line-like feature at about 0.81~keV; this could be due to the Fe-L line complex from the hot plasma (the basic {\em vpshock} model in {\em XSPEC} does not completely model the Fe-L transitions).  Due to the large number of events in this spectrum, its reduced $\chi^{2}$ (Table~\ref{tbl:diffuse_spectroscopy}) is large, although both the fit and fit residuals in Figure~\ref{fig:spectra}a seem reasonable and comparable to the global fits of other star-forming regions shown in Figure~\ref{fig:spectra}, below 2~keV at least.    

The NEI components' absorbing columns and plasma temperatures derived from this global spectrum are quite similar to the values obtained for individual tessellates in \citet{Townsley11b}, especially when only the ``inside'' tessellates (those contained within the global extraction region outlined in green in Figure~\ref{fig:carina+m17image}) are considered.  The first two NEI components (kT1 and kT2) have similar temperatures but different density-weighted ionization timescales; this may indicate a single plasma component in the process of transitioning from non-equilibrium to equilibrium, or it may show that kT2 is a lower-density plasma than kT1.

The unresolved pre-MS star components (cyan curves) have temperatures of 0.9~keV and 2~keV and a minimal absorbing column ($N_{H} = 0.1 \times 10^{22}$~cm$^{-2}$), possibly indicating that foreground field stars as well as pre-MS stars in Carina make up this unresolved stellar population.  The low normalization of these cyan curves (in Figure~\ref{fig:spectra}a) compared to the NEI components indicates that unresolved stellar emission contributes minimally to Carina's diffuse emission, as \citet{Townsley11b} concluded.  Specifically, the intrinsic total-band (0.5--7~keV) surface brightness of this component (both cyan curves combined) is $SB_{pre-MS} = 5.1 \times 10^{30}$~erg~s$^{-1}$~pc$^{-2}$ and its intrinsic total-band luminosity is $L_{pre-MS} = 6.1 \times 10^{33}$~erg~s$^{-1}$.  This is 3.6\% of the global diffuse luminosity for the CCCP (reported below) and is consistent with the estimates in \citet{Townsley11b} that were obtained using different methods.  Thus we conclude that the spectral model used here to estimate the contribution from unresolved stars to the global diffuse emission in these GHIIR's is reasonable and we will use the same method on the other targets in this study.

The spectral fit to Carina's global diffuse emission is significantly improved by the addition of gaussian lines at energies where no lines appear in the NEI plasmas (no matter what abundance enhancements are employed).  In \citet{Townsley11b} we proposed that these emission lines could be due to charge exchange when the hot plasma filling Carina's evacuated cavities interacts with the many cold surfaces distributed throughout the complex in the form of molecular clouds, ridges, pillars, and clumps (see that paper and \citealt{Townsley11a} for multiwavelength images of the Carina complex).  In this integrated global spectrum, many such lines are seen; their energies and plausible elemental origins (based on the emission line energies seen in hot plasmas) are:  0.64~keV (O), 0.77~keV (O or Fe), 1.07~keV (Ne), 1.34~keV (Mg), 1.54~keV (Mg), 1.80~keV (Si), 1.98~keV (Si), 2.61~keV (S), and 6.50 (Fe).  Enhanced abundances of O, Ne, Si, S, and Fe were needed in the thermal plasma models to account for other line features in the spectrum.  Since the absorbing columns to Carina's NEI plasmas are low, we did not include an absorption component for these gaussian lines, thus their fluxes are lower limits.  We could have achieved a better goodness-of-fit by adding more gaussian lines to the Carina global spectrum, especially above 2~keV, but we chose not to add any more complexity to this already complicated fit.  

We note that the spatially-resolved spectral modeling in the CCCP diffuse emission paper showed that a variety of gaussian line energies were required to model Carina's diffuse tessellates, with different lines required for different tessellates.  If these are charge exchange lines, this behavior implies that a range of physical conditions exists across the Carina Nebula, perhaps including differences in plasma abundances and densities, neutral material abundances and densities, and shock speeds as the hot plasma impinges on the cold neutral material.  The global Carina spectrum presented here thus represents, not surprisingly, a complex amalgam of physical conditions integrated across the entire Nebula.  The global spectra of the other GHIIRs presented below similarly represent integrations of a range of complicated physical interactions.

The last 4 columns in Table~\ref{tbl:diffuse_spectroscopy} give the intrinsic surface brightness (luminosity per square parsec) of the 3 NEI plasmas and the summed gaussian lines.  Summing these values for Carina, we see that the total-band intrinsic luminosity of Carina's diffuse emission (counting only the central region of the CCCP defined in Figure~\ref{fig:carina+m17image}) is $L_{tc} = 1.71 \times 10^{35}$~erg~s$^{-1}$; 57\% of that luminosity comes from the first NEI component (kT1) and less than 7\% comes from the gaussian lines in the spectral model.

\subsection{M17}


All 3 NEI plasma components contribute substantially to M17's global diffuse emission, with kT3 providing 56\% of the intrinsic diffuse luminosity.  Its intermediate density-weighted timescale implies that kT3 may be marginally in non-equilibrium and/or that it may be fairly low-density.  The 3 diffuse plasmas exhibit substantially different absorbing columns, with kT3 suffering 6 times more obscuration than kT1.  Such global extinction values are hard to interpret, since it is clear from the multiwavelength morphology of M17 that the obscuration changes across the field.  More detailed studies of the tessellated diffuse emission will allow us to make maps of the obscuration towards each NEI component; see \citet{Townsley11b} for examples.

The first 2 NEI components have the same plasma temperature but very different ionization timescales.  This perhaps indicates shocked gas that is in the process of returning to equilibrium but has not yet fully reached it; again, different densities could be an alternate explanation.  Although our data do not provide strong constraints on these ionization timescales, the very different timescale values for kT1 and kT2 are distinctive.  Since kT2 also has high emission measure, accounting for 35\% of M17's diffuse luminosity, there appears to be a substantial shock process at work here.  We will consider the nature of this shock process in Section~\ref{sec:discussion}, where we compare the global diffuse X-ray spectrum of M17 to other massive star-forming regions.

Unresolved pre-MS stars are modeled with a high absorbing column ($N_{H} = 1.2 \times 10^{22}$~cm$^{-2}$) similar to that found for kT3.  This implies that unresolved foreground stars do not contribute substantially to this contaminating population; it is reasonable that foreground star contamination would be very much reduced for this and the other GHIIRs in this study compared to Carina, since the CCCP was a large-area survey with 22 ACIS-I pointings and the other datasets consist of just 1 or 2 ACIS-I pointings.  The level of unresolved pre-MS star contamination in M17 does not appear to be high, according to our spectral fits; the thermal plasma components (cyan curves) have temperatures of 0.9~keV and 2.2~keV, with a combined intrinsic total-band surface brightness of $SB_{pre-MS} = 1.6 \times 10^{31}$~erg~s$^{-1}$~pc$^{-2}$ and intrinsic total-band luminosity of $L_{pre-MS} = 1.4 \times 10^{33}$~erg~s$^{-1}$ (6.8\% of the global diffuse luminosity for M17 given below).  This is probably because our \Chandra observation of M17's main ionizing cluster is long, $>$300~ks, so the point source detection sensitivity is relatively good.  Additionally, the hot plasma outflow from these massive stars is seen roughly edge-on, flowing through a ``crevice'' in the GMC towards the east, so it is not superposed on the cluster across most of the field (see Figure~\ref{fig:m17image}). 

As in Carina, M17's global diffuse X-ray spectrum requires the addition of gaussian lines to achieve a reasonable spectral fit.  As we concluded for Carina, charge exchange at the many hot/cold interfaces sampled by our \Chandra coverage of M17 is a plausible explanation for these spectral features.  The line energies and plausible elemental origins are:  0.65~keV (O), 0.79~keV (O or Fe), 0.90~keV (Ne), 1.30~keV (Mg), 1.95~keV (Si), 2.25~keV (S), 3.16 (S), and 6.43 (Fe).  Enhanced abundances of Ne, Si, and S were needed in the thermal plasma models to account for other line features in the spectrum.  The total-band intrinsic luminosity of M17's diffuse emission is $L_{tc} = 2.0 \times 10^{34}$~erg~s$^{-1}$, with just 3\% of that coming from the gaussian lines.

\subsection{NGC~3576}

As described above, this embedded GHIIR and the older stellar cluster seen to its north exhibit very different X-ray spectra.  Figure~\ref{fig:spectra}c shows the spectrum of the northern extraction region that was shown in Figure~\ref{fig:ngc3576image}; it includes a hard component that will be explored below.  In contrast, the spectrum of the southern extraction region (Figure~\ref{fig:spectra}d) shows no such hard component.

\subsubsection{The Southern Soft Outflow}

The bright, soft X-ray emission seen in the southeast part of the ACIS observation of NGC~3576 (Figure~\ref{fig:ngc3576image}) was a surprise, because previous studies of the ionizing massive stellar cluster showed that it was still enshrouded in its natal GMC.  The X-ray spectrum of this southeast extension shows little emission above 2~keV mainly because its extraction region is small compared to the other global spectra studied here, so little background emission is sampled.

This spectrum is situated below M17's spectrum in Figure~\ref{fig:spectra} because its shape is quite similar to M17's spectrum below 1.5~keV.  Given the morphological similarity of this soft X-ray emission to the X-ray outflow in M17, it is easy to imagine that this is also an outflow from its GHIIR.  This might be the hot plasma generated by NGC~3576's obscured massive stellar cluster that has found its way through a ``crevice,'' a lower-density pathway at the edge of the GMC, and is just now emerging from the front side of that cloud where we can detect it.  In this scenario, it is analogous to M17's hot plasma outflow, but viewed at a more face-on angle.

This southern soft emission from NGC~3576 is dominated by the kT1 plasma component with a temperature of 0.3~keV, with a small contribution from a higher-temperature component (kT2) that appears to be far from equilibrium or lower-density, given its short ionization timescale.  The dominant component is quite highly absorbed; this might be expected if our scenario of hot plasma just finding its way out of its confining cloud is correct.  The kT2 component is less absorbed and may indicate that this plasma encounters mild shocks as it makes its way out of the GMC.  No model component for unresolved pre-MS stars was needed to model this spectrum.  This is reasonable, since this outflow appears far from the NGC~3576 obscured cluster and even farther from the northern revealed cluster.

A gaussian at 0.72~keV is necessary for a good spectral fit; it may represent charge exchange emission from the many cold surfaces that surround this narrow outflow as it emerges from its GMC.  
Its energy is most consistent with oxygen emission but it could be an Fe-L line.  Enhanced abundances of O and Ne were needed in the thermal plasma models to account for other line features in the spectrum.  The total-band intrinsic luminosity of NGC~3576's southern outflow is $L_{tc} = 1.1 \times 10^{34}$~erg~s$^{-1}$, with 16\% of that coming from this gaussian line.  This high fraction of what might be charge exchange emission makes sense given that this plasma outflow encounters many cold interfaces as it emerges from its crevice in the GMC.

\subsubsection{The Northern Cavity Region}

The X-ray emission in the apparent IR cavity north of the obscured GHIIR NGC~3576 (Figure~\ref{fig:ngc3576image}) is distinctive from other large-scale emission studied here in that a power law component is required to achive a good spectral fit (see Figure~\ref{fig:spectra}).  This is perhaps not unexpected, since the field contains the pulsar PSR~J1112-6103 and we detect both the pulsar and its pulsar wind nebula in this ACIS observation (described above).  The hard, perhaps non-thermal X-ray emission modeled by the power law in our global X-ray spectrum of this region appears to fill the IR cavity; it is likely the signature of a cavity supernova associated with the NGC~3576 OB Association.  This power law component has a luminosity of $L_{tc} = 3.0 \times 10^{33}$~erg~s$^{-1}$, 24\% of the total X-ray luminosity in this extraction region ($L_{tc} = 1.2 \times 10^{34}$~erg~s$^{-1}$).  

The 3 soft NEI plasma components in our canonical model are all needed to fit the NGC~3576 northern cavity spectrum (see Table~\ref{tbl:diffuse_spectroscopy}).  The first two components (kT1 and kT2) have the same absorbing column but quite different temperatures (0.5 and 0.1~keV respectively); kT2 has a short ionization timescale.  The third NEI component (kT3) is harder (0.7~keV) with a higher absorbing column.  It accounts for 48\% of the diffuse luminosity, with $L_{tc} = 5.9 \times 10^{33}$~erg~s$^{-1}$.

Unresolved pre-MS stars (cyan components) contribute modestly to the global spectrum, with plasma temperatures of 0.9~keV and 2.5~keV and a minimal absorbing column ($N_{H} = 0.2 \times 10^{22}$~cm$^{-2}$).  Their intrinsic total-band surface brightness is $SB_{pre-MS} = 2.0 \times 10^{30}$~erg~s$^{-1}$~pc$^{-2}$ and intrinsic total-band luminosity is $L_{pre-MS} = 3.5 \times 10^{32}$~erg~s$^{-1}$, 2.9\% of the global diffuse luminosity for NGC~3576~North or 3.8\% of the remaining global diffuse luminosity after neglecting the power law component.

No gaussian lines are required to obtain an adequate fit to this spectrum.  The fit residuals show a possible line at $\sim$2.7~keV, but no lower-energy lines are obvious in the residuals.  This is perhaps an indication that charge exchange is not an important X-ray emission mechanism for this region.  The extraction region used for this spectrum sits mostly inside the IR cavity outlined by {\em MSX} data (Figure~\ref{fig:cavity}), thus it is likely not sampling as many hot/cold interfaces as the global spectra from other star-forming regions shown here.  This lack of charge exchange lines may imply that few cold clumps remain inside this cavity, perhaps consistent with our speculation that the stellar cluster found here is a little older than those in the other star-forming regions we are considering.


\subsection{NGC~3603}

\subsubsection{The Main Complex}

The spectrum extracted for the main NGC~3603 region covered all of the ACIS-I pointing except a section on the west side (see Figure~\ref{fig:ngc3603image}) that appeared to be spatially distinct because it sits at the eastern edge of the IR cavity north of NGC~3576 shown in Figure~\ref{fig:cavity}.  Our spectral analysis excluded an area around the center of the massive stellar cluster in NGC~3603 because the short \Chandra dataset is certainly dominated by unresolved point sources near the cluster core.  

The diffuse X-ray spectrum of the main NGC~3603 complex is dominated by the kT1 component, which has a relatively high temperature (0.5~keV) and a short ionization timescale implying that it is indeed a plasma in a non-equilibrium state.  The second NEI component (kT2) is softer and minimally-absorbed; it may be a foreground component unrelated to the distant NGC~3603 complex.  Even if it is associated with NGC~3603, it contributes minimally to the diffuse emission X-ray luminosity.  The kT3 component has a temperature and absorption consistent with kT1 but with a long ionization timescale.  Again we may be seeing a single dominant plasma in the process of transitioning from non-equilibrium to equilibrium ionization.  In this case, the non-equilibrium component (kT1) dominates the luminosity, contributing 86\% of the total $L_{tc} = 2.6 \times 10^{35}$~erg~s$^{-1}$.

The cyan lines in the spectral fit (Figure~\ref{fig:spectra}) again represent the unresolved pre-MS stars, with thermal plasma temperatures of 0.9~keV and 2.2~keV, total-band intrinsic surface brightness of $SB_{pre-MS} = 3.2 \times 10^{31}$~erg~s$^{-1}$~pc$^{-2}$, and total-band intrinsic luminosity of $L_{pre-MS} = 3.1 \times 10^{34}$~erg~s$^{-1}$ (11.8\% of the global diffuse luminosity for NGC~3603 given above).  We found that this component had the same absorbing column as the diffuse components kT1 and kT3, so we believe that these are mostly unresolved stars in the NGC~3603 complex.  Their emission dominates the X-ray spectrum above 1.5~keV; this unresolved pre-MS star contamination is much stronger in NGC~3603 than in the other star-forming regions described here (and would have been worse if we had not excluded the central cluster from consideration; recall the mask in Figure~\ref{fig:ngc3603image}).  This is not surprising, given that this is one of the most massive young clusters in the Galaxy, it is comparatively far away, and our \Chandra observation was quite short, so our point source detection sensitivity was very shallow.  We have just obtained new \Chandra observations of this target that are 10 times deeper, so future analysis of NGC~3603's diffuse emission will be less contaminated by point sources.

Even with this strong contamination, NGC~3603's diffuse X-ray emission is spectrally distinct enough that we were able to characterize its properties.  The new \Chandra observations will allow us to examine its spatial variations as well.  Additionally, we will try to establish whether any of this emission is due to the foreground NGC~3576 star-forming complex, superposed on NGC~3603's line of sight.

No elevated plasma abundances or gaussian lines were necessary to obtain an acceptable fit to this spectrum, although the residuals show structure at 2, 3, and 6~keV that may be due to Si, S, and Fe line emission.  Given the GMC interfaces that surround NGC~3603 and that are captured by our ACIS-I pointing, we might have expected to see charge exchange emission lines here as we saw in other star-forming regions.  Again the longer \Chandra observation may be needed to reveal such features.

\subsubsection{The Western Diffuse Emission}

At the western edge of the NGC~3603 observation, we isolated the diffuse emission for separate consideration from the main NGC~3603 star-forming region, since it could be due to the foreground NGC~3576 complex.  The spectral fit of this western emission, though, reveals emission components that are more consistent with the rest of NGC~3603 than they are with NGC~3576.  Again kT1 and kT3 suffer large absorbing columns and have comparable temperatures, while kT2 has much lower obscuration and a lower temperature.  The ionization timescales for kT1 and kT3 again imply an NEI plasma transitioning to equilibrium or that kT1 is a much lower-density plasma than kT3.  As for the main NGC~3603 spectrum, kT1 dominates the luminosity of the region, accounting for 85\% of the total $L_{tc} = 1.2 \times 10^{35}$~erg~s$^{-1}$; thus the plasma is primarily in a low-density or non-equilibrium state.  This luminosity is remarkably high, given that the region sampled here is only 22\% as big as the main NGC~3603 extraction region.  

This western extraction is far from the center of the massive cluster in NGC~3603, thus it is only minimally affected by unresolved pre-MS stars.  The pre-MS model component requires only $A_{V} \sim 1$~mag of extinction, implying that it is probably composed primarily of stars in the closer star-forming region NGC~3576 and/or foreground field stars.  It has thermal plasma temperatures of 0.8~keV and 2.2~keV, total-band intrinsic surface brightness of $SB_{pre-MS} = 1.8 \times 10^{30}$~erg~s$^{-1}$~pc$^{-2}$, and total-band intrinsic luminosity of $L_{pre-MS} = 3.9 \times 10^{32}$~erg~s$^{-1}$ (just 0.3\% of the global diffuse luminosity for NGC~3603~West given above).  As for NGC~3576~South, this extraction region is not centered on a star cluster, so the contamination from unresolved pre-MS stars is minimal.

This region also shows signs of charge exchange emission; gaussians at 0.76~keV, 1.44~keV, and 2.01~keV (possibly from O, Mg, and Si, respectively) were required to achieve an acceptable spectral fit, although only 4\% of the diffuse luminosity comes from these gaussian components.  These gaussians required an absorbing column similar to that found for kT1, so it appears that the charge exchange, as well as the hot plasma emission, is associated with the NGC~3603 complex at 7~kpc rather than with the foreground NGC~3576 complex at 2.8~kpc.  Since this western emission has X-ray plasma properties consistent with NGC~3603 but appears morphologically to be shadowed by the edges of the IR cavity north of NGC~3576, we are still faced with confusion regarding the distance of the cold absorbing components that thread around both NGC~3603 and NGC~3576.  Perhaps distant NGC~3603 diffuse X-ray emission is being shadowed by foreground structures in a spiral arm less than half as far away.  Only detailed velocity mapping on large spatial scales across these two star-forming complexes will disentangle these unfortunately-placed GHIIRs.

\subsection{30~Doradus}

\citet{Townsley06a} characterized 30~Dor's global diffuse emission using a background-subtracted spectrum with a celestial background sampled from the edges of the ACIS-I field.  This method is appropriate for 30~Dor because the diffuse X-ray emission does not fill the field, or at least it is very much fainter in the regions chosen as background samples than in the rest of the field.  As described above, we were unable to apply this technique to the Galactic star-forming regions, thus our spectral modeling of those regions had to be more complicated to include celestial background components.

We used the same 30~Dor spectral extraction here as was used in \citet{Townsley06a} but we re-fit the spectrum using NEI plasma components and included the possibility of gaussian lines due to CE.  Fit results are shown in Figure~\ref{fig:spectra}g and h; these spectra are clearly simpler in form than those of the Galactic regions described above, primarily because virtually all of the hard part of the spectrum went away in the celestial background subtraction.  In Figure~\ref{fig:spectra}g we assumed solar abundances for all elements in the {\em vpshock} model, while in Figure~\ref{fig:spectra}h we assumed 0.4$\times$ solar abundances for all elements, in accordance with the subsolar abundances found for the LMC \citep{Russell90}.  Fits were performed with the abundances linked together between the 3 NEI plasma components; abundances were frozen to these values for the initial fits, then each element's abundance was allowed to vary to higher values if such higher values improved the fits.

Both fits are quite similar; these data do not give a clear indication of which family of model abundances is better for 30~Dor.  All model components were found to require very similar absorbing columns, so we simplified the fits by linking all of the absorption components to that of kT1.  The second NEI component (kT2) was not needed for these fits.  Neither kT1 nor kT3 required a short ionization timescale, implying that the plasmas are in (or near) ionization equilibrium.  

Both spectral models require a gaussian at 0.75~keV to achieve a good fit.  The residuals imply that a second gaussian at $\sim$0.65~keV might also be warranted, but attempts to include such a line did not improve the fit.  As we found above, these gaussians might be an indication of charge exchange emission, in this case probably due to oxygen.  Enhanced abundances of O, Ne, Mg, and (in the solar abundance fit) Fe were needed in the thermal plasma models to account for other line features in the spectrum.

No pre-MS star model components were necessary to fit the 30~Dor diffuse X-ray spectrum, even though the field should be filled with hundreds of thousands of unresolved pre-MS stars, given the long history of vigorous star formation in this complex.  The simplicity of our spectral model for 30~Dor must reflect the very shallow sensitivity of this short \Chandra observation; either the integrated emission of 30~Dor's young stellar population is too faint to be detected or it was subtracted away in our celestial background subtraction.  The latter explanation would require that the background regions have a similar surface brightness in unresolved young stars as the central part of 30~Dor.  Perhaps this latter explanation is not impossible, given the distributed population of young stars that we found in Carina \citep{Townsley11a,Feigelson11}.

The intrinsic luminosities of 30~Dor's diffuse emission that are inferred from our spectral fits are quite high due to the soft plasma temperature of kT1 and its fairly substantial absorbing column ($N_{H1}$).  For the model that used solar abundances, $L_{tc} = 1.87 \times 10^{37}$~erg~s$^{-1}$ with 57\% coming from kT1; for the model with subsolar abundances, $L_{tc} = 1.23 \times 10^{37}$~erg~s$^{-1}$ with 40\% coming from kT1.  For both models, 6\% of the luminosity comes from the gaussian line at 0.75~keV, again possibly due to charge exchange with neutral atoms in 30~Dor's ISM.  We expect that longer \Chandra observations of 30~Dor will yield much richer spectral information and that spatially-resolved X-ray spectral analysis across this bright, complex structure is necessary to understand the wide range of physical processes at work here.


    
\section{DISCUSSION \label{sec:discussion}}

\subsection{Integrated X-ray Emission from GHIIRs}

For convenience, we summarize total values for the intrinsic diffuse X-ray emission of our GHIIRs in Table~\ref{tbl:global}.  These regions are compared to each other in Section~\ref{sec:carinacomp} below.  This summary table will be useful for comparing the diffuse X-ray emission in resolved GHIIRs to more distant, unresolved complexes as well as to smaller Galactic star-forming regions.

\setlength{\tabcolsep}{0.03in}
\renewcommand{\arraystretch}{1.0}
\begin{deluxetable}{@{}lccl@{}}
\centering \tabletypesize{\small} \tablewidth{7in}

\tablecaption{Integrated Global Diffuse Emission Properties
\label{tbl:global}}

\tablehead{
\colhead{Target} & \colhead{SB$_{tc}$} & \colhead{$L_{tc}$} & \colhead{L$_{gau,\Gamma}$} \\
\colhead{} & \colhead{($\times 10^{32}$ erg s$^{-1}$ pc$^{-2}$)} & \colhead{($\times 10^{34}$ erg s$^{-1}$)} & \colhead{} \\

\numberthecolumn & \numberthecolumn & \numberthecolumn & \numberthecolumn 
\setcounter{column_number}{1}
}
\startdata
Carina          & 1.4 &     \phd\phd17.1      & \phd7\% \\
M17             & 2.3 &  \phd\phd\phd2.0      & \phd3\% \\
NGC~3576~S      & 7.1 &  \phd\phd\phd1.1      & 16\% \\                            
NGC~3576~N      & 0.7 &  \phd\phd\phd1.2      & 24\% ($\Gamma$) \\
NGC~3603        & 2.7 &     \phd\phd26.0      & \nodata \\
NGC~3603~W      & 5.5 &     \phd\phd11.7      & \phd4\% \\
30~Dor solar+   & 5.6 &           1870.\phd   & \phd6\% \\
30~Dor subsolar & 3.7 &           1230.\phd   & \phd6\% 
\enddata

\tablecomments{
Col.\ (1):  Target name.
\\Col.\ (2): Absorption-corrected total-band surface brightness (intrinsic luminosity per unit area) of the diffuse emission (the sum of Columns 20-23 in Table~\ref{tbl:diffuse_spectroscopy}).
\\Col.\ (3): Absorption-corrected total-band luminosity of the diffuse emission (Col.\ 2 multiplied by the area in square parsecs given in Column 3 of Table~\ref{tbl:diffuse_spectroscopy}).
\\Col.\ (4): Fraction of Col.\ (3) attributed to gaussian lines or to a power law component (for NGC~3576 North).
}          

\end{deluxetable}

For extragalactic studies, it would also be helpful to be able to estimate what fraction of the X-ray luminosity of an unresolved star-forming region is due to point sources and what fraction is due to diffuse emission.  Unfortunately, given the variety of X-ray-emitting objects that might inhabit a GHIIR, a reliable generalization of the relative contributions of point-like and diffuse components in such a region is not possible.  For example, in Carina, the X-ray emission from \etacar dominates over all other individual point sources, and sometimes over whole integrated populations, if it is observed during its active phase \citep{Corcoran05}.  In 30~Dor, the young supernova remnant N157B (and its associated pulsar and pulsar wind nebula) generates $\sim$20\% of the X-ray luminosity of the whole complex \citep{Townsley06a}; we find a similar example here with NGC~3576OB (NGC~3576~N in Table~\ref{tbl:global}). 

Thus the fractional contribution to the total X-ray luminosity of a GHIIR from point sources or from diffuse emission can be dominated by one or a few spectacular and unusual sources.  This will get worse as the star-forming complex ages and supernovae and X-ray binaries start to dominate the X-ray emission.  It might be possible to limit study to an energy range (perhaps 0.5--0.7~keV for ACIS-I) that minimizes the contribution from point sources, but this also limits the available counts and increases the need for accurate calibration (e.g., of the ACIS Optical Blocking Filter contamination, which strongly affects the ACIS-I throughput below 1~keV).  Even such an energy restriction will not prevent the X-ray luminosity of an unresolved star-forming complex from being dominated by a recent supernova.

\subsection{Comparing Carina to Other Star-forming Regions \label{sec:carinacomp}}

For this study, the most distinctive aspect of our \Chandra data on the Carina Nebula is the large number of counts that go into Carina's global diffuse X-ray spectrum.  This allows us to see faint features, especially faint charge exchange lines, that are not discernable in the lower-sensitivity spectra we have for other targets.  The basic spectral model form that we developed for the CCCP \citep{Townsley11b} is quite successful in modeling the global diffuse X-ray emission from other star-forming regions; this must be due at least in part to its complexity.  

\subsubsection{M17}


The spectrum most similar to Carina's is M17's, partly because it contains a comparatively large number of counts so the charge exchange lines are necessary for a good fit.  The NEI components are similar as well, though.  Both targets contain a soft plasma (kT$\sim$0.3~keV) with a mix of NEI timescales that implies a range of densities and/or a plasma transitioning from NEI to CIE.  Both targets also contain a harder plasma (kT$\sim$0.6~keV); this harder component appears to be in CIE in Carina but not in M17, where it also dominates the X-ray luminosity in the diffuse emission.  Both targets require enhanced plasma abundances and gaussians to model the wide array of line-like features in the spectra.  Notably, the energies of the gaussian lines found for Carina are often quite similar to those found for M17, implying that these features are not random spectral fluctuations.

While there is evidence for supernova activity in Carina \citep[e.g.,][]{Hamaguchi09,Townsley11a,Townsley11b} that might explain its bright diffuse X-ray emission and NEI plasmas, M17's ionizing cluster NGC~6618 is very young \citep{Hanson97} so no such activity is expected.  We found M17's global diffuse intrinsic X-ray luminosity to be a factor of 8.5 times fainter than Carina's, but its average intrinsic surface brightness is actually higher than Carina's, as is the surface brightness for the strongly-NEI component (kT2).  Are M17's massive star winds strong enough to support NEI plasma emission brighter than that seen on average in Carina, perhaps due to plasma confinement in the ``crevice''?  Or could M17's diffuse X-ray emission be so unusually bright because that complex also has experienced supernova activity?  There is evidence for an older massive cluster in a bubble structure north of M17 that may have influenced the formation of today's massive obscured cluster \citep{Povich09} and there are OB stars spread across the field covered by our \Chandra data that are not concentrated in the NGC~6618 cluster \citep{Ogura76}.  Perhaps this older population could be contributing cavity supernova emission to the \Chandra data.  We will explore this question further in a future detailed study of M17's diffuse X-ray emission.

\subsubsection{NGC~3576}

The diffuse X-ray emission associated with the NGC~3576 OB Association north of the embedded GHIIR has an average surface brightness about half that of Carina.  It also has an NEI plasma component (kT2) with a short ionization timescale, but the plasma temperature of this component is very soft (0.1~keV).  Of course the most prominent spectral differences here are the non-thermal emission in NGC~3576's northern cavity and the lack of evidence for charge exchange.  If the non-thermal emission is indeed due to a cavity supernova remnant, it is likely that this supernova exploded more recently than any such event in Carina, which shows no non-thermal X-ray emission.  Recent supernova activity could also explain the presence of the pulsar PSR~J1112-6103 along the line of sight to this cavity, which again is thought to be young \citep[$<10^{5}$~yr,][]{Manchester01} (although the large-scale non-thermal emission is not necessarily from the same supernova that resulted in the pulsar).  Based on the presence of this non-thermal X-ray emission and the supernova activity that it implies, then, we conjecture that PSR~J1112-6103 is physically associated with the NGC~3576 OB Association and the young X-ray-emitting stellar cluster that we found there (called NGC~3576OB below).

We can compare NGC~3576OB to Trumpler~15 (Tr15) in Carina, thought to be the oldest ($\sim$8~Myr) prominent stellar cluster in that complex \citep{Dias02,Tapia03}.  Using the CCCP data, \citet{Wang11} show that Tr15 possesses a rich low-mass pre-MS stellar population but appears deficient in massive stars, perhaps because they have already exploded as supernovae, enriching Carina's ISM and adding to its diffuse X-ray emission.  The neutron star recently discovered in Carina \citep{Hamaguchi09} could have come from Tr15, analogous to (although older than) the pulsar in NGC~3576OB.  

The young, massive cluster powering the NGC~3576 GHIIR sits just inside the edge of a GMC; it might have been triggered by feedback from NGC~3576OB \citep{Garcia94,Girardi97}.  It sits about 15$\arcmin$ south of that cluster ($\sim$12~pc at a distance of 2.8~kpc).  Perhaps Tr14 is the Carina analogy.  Tr14 is probably slightly older than the embedded NGC~3576 cluster since Tr14 has already emerged from its GMC.  Tr14 is roughly the same distance ($\sim$8~pc) from Tr15 as we see between the two clusters in NGC~3576.  The CCCP showed that there is a substantial bridge of enhanced stellar density between Tr14 and Tr15 \citep{Feigelson11,Wang11}; a similar structure may exist between NGC~3576OB and the embedded GHIIR (Figure~\ref{fig:ngc3576image}), but better \Chandra spatial coverage of the complex would be necessary to establish this with certainty.

Nothing resembling NGC~3576's H$\alpha$ loops (Figure~\ref{fig:cavity}) is seen in Carina or any other target studied here.  This loop morphology implies that the embedded GHIIR's outflow into the cavity \citep{dePree99} is plowing up material; that material either started out ionized or was ionized by NGC~3576OB.  The H$\alpha$ loops must be indicating where the swept-up material reaches its highest density.  Perhaps no similar structures are seen in Carina or in other star-forming regions because they possess too much turbulence or overlapping outflows from young stars; such large-scale loops might not last long in a place like Carina, where mass loading from its many cold clumps, pillars, and ridges, and outflows from many directions \citep{Smith10}, leave few quiescent cavities.  The organized outflow from the NGC~3576 embedded cluster may stay intact because NGC~3576OB has largely cleared its cavity of such cold material and its stars are old enough that it also lacks strong, competing outflows.  The major mystery for this scenario is the pulsar, though; it seems unlikely that the tenuous H$\alpha$ loops would not be destroyed by a supernova exploding inside the cavity into which they were expanding.

We said above that Tr14 appears to have a soft bright outflow to the south that is reminiscent of M17; NGC 3576's soft X-ray outflow to the southeast of the embedded cluster is even more like M17's, seen more face-on.  It has spectral characteristics ($kT \sim 0.3$~keV) similar to the M17 outflow; this appears to be another X-ray outflow caused by the powerful winds of massive stars, seen here in an earlier phase than in M17, just now breaking out of the front edge of the cluster's natal GMC.  Surprisingly, its intrinsic luminosity is as high as what we found for the much larger cavity to the north and is half the value for M17.  Since it covers such a small spatial region, though, its average intrinsic surface brightness is much larger than any of the regions we have studied so far.  This is perhaps explained by geometry; if this is a powerful confined outflow from the massive embedded cluster powering the GHIIR but now seen nearly straight into the crevice along the long dimension of the outflow, its emission is concentrated into a small area, resulting in a high surface brightness.

A gaussian line at 0.72~keV must be added to the spectral model for the southern outflow to achieve a good fit; this is close in energy to the most prominent gaussian needed for Carina as well.  While we suggested that no gaussian lines were needed to model NGC~3576's northern cavity emission because few hot/cold interfaces generating charge exchange were sampled there, seeing the line again (quite strongly) in the southern outflow might well be expected, since this ACIS extraction region samples many of these hot/cold interfaces as the hot plasma makes its way through the crevice in the GMC.

\subsubsection{NGC~3603}

NGC~3603's diffuse X-ray emission appears faint (Figure~\ref{fig:ngc3603image}) due to the shallow sensitivity of our \Chandra observation and the relatively large intervening absorption to this distant GHIIR.  The spectral model for the main NGC~3603 region (the full ACIS-I pointing shown in Figure~\ref{fig:ngc3603image} less the western region) shown in Figure~\ref{fig:spectra} reveals substantial contamination by unresolved pre-MS stars in this massive monolithic cluster.  This contamination was expected and helped to motivate a new, longer \Chandra observation of this target, just recently obtained (October 2010).  This new dataset will resolve out several thousand stars in the ACIS-I field, enabling a cleaner study of NGC~3603's diffuse emission.

In the meantime, our spectral analysis of the original ACIS-I dataset did allow us to characterize NGC~3603's diffuse emission and we find that its faint appearance is misleading.  The plasma temperatures are similar to what we found in Carina, but for NGC~3603 the hotter, low-timescale component (kT1 = 0.5~keV) accounts for 86\% of the total X-ray luminosity.  This total intrinsic luminosity is strikingly high ($L_{tc} = 2.6 \times 10^{35}$~erg~s$^{-1}$), a factor of 1.5 brighter than Carina's integrated diffuse emission but distributed over a smaller area (the average surface brightness is almost twice that in Carina).  

Recent data from {\em Hubble}/WFC3 \citep{Beccari10} give evidence for a stellar population in NGC~3603 older than 10~Myr; Carina may also contain such an older population \citep[][and references therein]{Townsley11a}.  The bright diffuse X-ray emission that we infer from our spectral fits to NGC~3603 and the presence of stellar populations several Myr old suggest that NGC~3603 may also have seen supernova activity.  Our upcoming study of NGC~3603's X-ray-emitting point source population will certainly include a search for neutron star candidates, as we did in Carina (Townsley et al.).

The western diffuse emission that we analyzed separately in the NGC~3603 ACIS-I data looks much like the main NGC~3603 emission.  The 3 NEI components in our spectral model for this emission show absorbing columns similar to those found for the main NGC~3603 field.  Plasma temperatures are similar but slightly softer than the main NGC~3603 field.  While the main NGC~3603 field spectral fit did not require the addition of gaussian components, they are necessary for the western field, contributing a similar fraction (4\%) to the total X-ray luminosity as seen in other GHIIRs.  Since the main NGC~3603 field should be sampling as many hot/cold interfaces as the western region and as the other GHIIRs studied here, the lack of charge exchange lines in that spectrum is puzzling.  Perhaps they are swamped by a combination of high obscuration and the strong unresolved pre-MS stars that dominate the X-ray spectrum above 1.5~keV.

Surprisingly, though, the western NGC~3603 emission has high luminosity ($L_{tc} = 1.2 \times 10^{35}$~erg~s$^{-1}$), with an average intrinsic surface brightness twice as high as the main NGC~3603 region and almost four times higher than Carina's average surface brightness.  Perhaps we are seeing a mix of distant X-ray emission from NGC~3603 and closer X-ray emission from the northern cavity of NGC~3576; if so, our X-ray luminosity and surface brightness would be over-estimated because we assumed that all X-ray emission was coming from the distance of NGC~3603.  Alternatively, perhaps some of the NGC~3576 northern cavity emission described above could in fact be coming from NGC~3603; the kT3 component in the NGC~3576 northern spectral fit had the same high absorbing column and similar temperature to kT3 in the NGC~3603 western fit.  Again, more detailed, spatially-resolved X-ray spectral fitting of the diffuse emission in these overlapping GHIIRs will be necessary to pursue these questions further.

\subsubsection{30~Doradus}

A detailed study of the diffuse X-ray emission in 30~Dor was presented in \citet{Townsley06a}, using the same short ACIS-I observation ($\sim$22~ks) described here.  Although much cruder methods were used there compared to our CCCP diffuse analysis of Carina \citep{Townsley11b}, there was already an indication that complex combinations of absorption, thermal plasma temperature, and intrinsic surface brightness were necessary to explain the spatial and spectral distributions of X-ray emission across the field.  The red outline of the CCCP survey superposed on 30~Dor's soft unresolved X-ray emission in Figure~\ref{fig:30dorimage} is a reminder that 30~Dor's diffuse X-ray emission is likely to be at least as complicated as what we saw in Carina and the other Galactic GHIIRs featured here, but the shallow sensitivity of current \Chandra observations limits the degree to which we can probe this complexity.  

Considering our current global diffuse X-ray emission spectral fits for 30~Doradus (Table~\ref{tbl:diffuse_spectroscopy}), we find a higher intrinsic luminosity than \citet{Townsley06a} found because we used an additional thermal plasma component here (kT1 = 0.2~keV) to achieve a better fit.  This is slightly softer than what we found for the kT1 component in Galactic GHIIRs; the higher-temperature component (kT3 = 0.55~keV) is consistent with the Galactic targets.  Both components show high ionization timescales implying equilibrium plasmas.  The NEI component that often showed short timescales for Galactic regions (kT2) is not needed to fit 30~Dor's global X-ray spectrum.  The average intrinsic surface brightness of 30~Dor's diffuse X-ray emission is comparable to that found for the brighter parts of Galactic GHIIRs; its total X-ray luminosity is 2--3 orders of magnitude higher than Galactic regions because the size of the complex is so much larger than Galactic GHIIRs.

In the CCCP introductory paper \citep{Townsley11a}, we compared the CCCP survey to the \Chandra Orion Ultradeep Project \citep[COUP,][]{Feigelson05}, an ACIS-I study of the Orion Nebula Cluster, the closest massive star-forming region.  We found that COUP was $\sim$430 times more sensitive than the CCCP, or that the CCCP was about the same as a 2~ks ACIS-I observation of the ONC.  We can apply similar arguments to compare the CCCP to the 22-ks ACIS-I observation of 30~Dor.  

30~Dor, at a distance of $\sim$50~kpc \citep{Feast99}, is 21.7 times farther away than Carina, at $\sim$2.3~kpc \citep{Smith06}. 30~Dor's angular coverage is then $\sim$14.5~pc~arcmin$^{-1}$ while Carina's is $\sim$0.7~pc~arcmin$^{-1}$.  The nominal CCCP integration time was 60~ks.  Since survey sensitivity goes as the integration time divided by the square of the distance, the CCCP is $\sim$1300 times more sensitive than the 30~Dor ACIS-I observation studied here.  Thus the 30~Dor ACIS-I observation is roughly equivalent to a 46-{\em second} ACIS-I observation of Carina; in that time we would have gathered $\sim$770 diffuse X-ray counts in our CCCP global extraction region and only two of the $>$14,000 CCCP X-ray point sources (\etacar and WR~25) would have been detected.  It is a testament to the size and power of 30~Dor that our 22-ks ACIS-I observation of that GHIIR gathered 32,400 diffuse counts \citep{Townsley06a} and detected over 150 massive stars and stellar systems \citep{Townsley06b}.

\section{CONCLUSIONS AND SUMMARY \label{sec:conclusions}}

Every GHIIR studied here appears to be the product of multiple generations of massive star formation, evolution, and feedback.  X-ray signatures of all of these processes are present in the complicated mix of point-like, unresolved, and truly diffuse X-ray emission that characterizes GHIIRs across the Milky Way and beyond.  Painstaking work using high-sensitivity observations with a high-spatial-resolution X-ray telescope is necessary to disentangle these many X-ray emission components.  Several important inferences emerge from this work.

\begin{itemize}
\item In the GHIIRs studied here, star formation plausibly has been underway for 10--100~Myr, thus the cumulative feedback of supernovae and fast stellar winds often makes it difficult to isolate the exact origin of the hot gas we observe in X-rays.
   
\item This hot gas generated by massive star feedback in a cluster flows to wherever the least resistance (pressure or density) is present in the surrounding ISM, thus diffuse X-ray emission is often detected at positions offset from the cluster(s).
   

\item For the targets and \Chandra observations examined here, contamination of the global diffuse emission spectrum by unresolved pre-MS stars, Galactic Ridge emission, or extragalactic sources is modest; such contamination becomes minimal when our extraction regions are not centered on a bright young stellar cluster.  For less extreme \hii regions with fainter diffuse X-ray emission, these contaminating components will play a bigger role and must be carefully modeled.

\item If similar star-forming regions are observed with insufficient angular resolution or sensitivity such that the X-ray-emitting point sources are not resolved, their contamination can dominate the total unresolved X-ray emission, depending on the energy range under consideration and the spatial distribution and evolutionary stage of the underlying stellar population.  Analysis of extragalactic GHIIRs can be complicated by these point source populations and by individual, highly-luminous X-ray sources that occasionally are found in young star-forming complexes.
  
\item The presence of strong X-ray emission lines in global spectra of GHIIRs that are not included in standard X-ray plasma emission models may be a result of the interaction between the hot gas and the cold molecular material that remains in these regions, giving rise to charge exchange.  The line at about 0.76~keV (possibly due to oxygen) is often the most prevalent, i.e., the first line to become necessary in the weak spectra from low-sensitivity observations.  More counts reveal more spectral structure, requiring more lines; those line energies are often close (but not identical) to the line energies regularly seen in hot plasmas due to abundance enhancements of certain common elements.  These extra lines have energies that are similar from target to target, implying that the process that generates them (be it charge exchange or some other emission mechanism) is common to \hii regions that exhibit bright diffuse X-ray emission.

\item If the lines that are not modeled by hot plasmas are in fact due to charge exchange at the hot/cold interfaces as we have postulated, we find that charge exchange is ubiquitous in massive star-forming regions, both in the Milky Way and beyond.  Thus GHIIRs are a new class of sources exhibiting charge exchange X-ray emission.

\item It is beyond the scope of this paper to try to model the physical conditions in GHIIRs to explain the range of plasma temperatures and variety of NEI versus CIE conditions that we measure, or to predict the charge exchange line strengths and line energies that should be present.  Future spatially-resolved, tessellated X-ray spectral fitting will likely provide a more appropriate testbed for such modeling than these global spectra offer, especially since our Carina study \citep{Townsley11b} showed that plasma temperatures, plasma abundances, and charge exchange lines vary between Carina tessellates. 

\end{itemize}

In summary, this experiment in fitting the global spectra of unresolved X-ray emission in GHIIRs with the spectral model developed for the CCCP study of the Carina Nebula has shown that Carina appears to be a typical example of a massive star-forming complex with bright diffuse X-ray emission, likely originating in a complex mix of emission sources:  OB wind shocks, cavity supernovae, mass loading, and/or charge exchange interactions between hot plasmas and the many cold surfaces that characterize these regions.

For the most part, the expectations for other GHIIRs that we developed by studying Carina were fulfilled:  ACIS observations that capture many hot/cold interfaces (M17, NGC~3576 South, NGC~3603 West, 30~Doradus) show spectral evidence for what appears to be charge exchange emission while those that do not image such interfaces (NGC~3576 North) do not.  The one exception is the main NGC~3603 field, where the ACIS pointing certainly captures these hot/cold interfaces but no charge exchange lines are seen; we have in hand a new \Chandra observation of this target that is 10 times longer than the dataset presented here, so we will soon learn if the absence of charge exchange lines seen here was simply due to sensitivity limitations or whether this region is truly physically distinct.

\acknowledgments
We very much appreciate the time and effort donated by our anonymous referee to improve this paper; the enthusiastic and detailed review was most helpful.  This work was supported by \Chandra X-ray Observatory grants GO8-9131X, GO8-9006X, AR9-0001X, GO6-7006X, GO5-6080X, and GO4-5007X (PI:  L.\ Townsley) and by the ACIS Instrument Team contract SV4-74018 (PI:  G.\ Garmire), issued by the \Chandra X-ray Center, which is operated by the Smithsonian Astrophysical Observatory for and on behalf of NASA under contract NAS8-03060.  We thank R{\'e}my Indebetouw for providing the IRAC image of NGC~3576 and Eric Feigelson for comments on an early draft.  This research made use of data products from the Midcourse Space Experiment, the NASA/IPAC Infrared Science Archive, and NASA's Astrophysics Data System.

{\it Facilities:} \facility{CXO (ACIS)}, \facility{SST (IRAC)}, \facility{MSX ()}.



\begin{thebibliography}

\bibitem[Albacete-Colombo et al.(2003)]{Albacete03} Albacete-Colombo, J.~F., M{\'e}ndez, M., \& Morrell, N.~I.\ 2003, \mnras, 346, 704 

\bibitem[Arnaud(1996)]{Arnaud96} Arnaud, K.~A.\ 1996, Astronomical Data Analysis Software \& Systems V (ASP Conf.\ Ser.\ 101), ed.\ G.~H.~Jacoby \& J.~Barnes (San Francisco, CA: ASP), 17

\bibitem[Barbosa et al.(2003)]{Barbosa03} Barbosa, C.~L., Damineli, A., Blum, R.~D., \& Conti, P.~S.\ 2003, \aj, 126, 2411 

\bibitem[Beccari et al.(2010)]{Beccari10} Beccari, G., et al.\ 2010, \apj, 720, 1108

\bibitem[Broos et al.(2007)]{Broos07} Broos, P.~S., Feigelson, E.~D., Townsley, L.~K., Getman, K.~V., Wang, J., Garmire, G.~P., Jiang, Z., \& Tsuboi, Y.\ 2007, \apjs, 169, 353

\bibitem[Broos et al.(2010)]{Broos10} Broos, P.~S., Townsley, L.~K., Feigelson, E.~D., Getman, K.~V., Bauer, F.~E., \& Garmire, G.~P.\ 2010, \apj, 714, 1582 

\bibitem[Broos et al.(2011)]{Broos11} Broos, P.~S., et al.\ 2011, \apjs, submitted (CCCP Catalog Paper)

\bibitem[Chini \& Hoffmeister(2008)]{Chini08} Chini, R., \& Hoffmeister, V.\ 2008, Handbook of Star Forming Regions, Volume II, 625 

\bibitem[Chu \& Mac Low(1990)]{Chu90} Chu, Y.-H., \& Mac Low, M.-M.\ 1990, \apj, 365, 510 

\bibitem[Churchwell et al.(2009)]{Churchwell09} Churchwell, E., et al.\ 2009, \pasp, 121, 213

\bibitem[Clark et al.(2008)]{Clark08} Clark, J.~S., Muno, M.~P., Negueruela, I., Dougherty, S.~M., Crowther, P.~A., Goodwin, S.~P., \& de Grijs, R.\ 2008, \aap, 477, 147

\bibitem[Corcoran(2005)]{Corcoran05} Corcoran, M.~F.\ 2005, \aj, 129, 2018 

\bibitem[de Pree et al.(1999)]{dePree99} de Pree, C.~G., Nysewander, M.~C., \& Goss, W.~M.\ 1999, \aj, 117, 2902

\bibitem[Dias et al.(2002)]{Dias02} Dias, W.~S., Alessi, B.~S., Moitinho, A., \& L{\'e}pine, J.~R.~D.\ 2002, \aap, 389, 871 

\bibitem[Ebeling et al.(2002)]{Ebeling02} Ebeling, H., Mullis, C.~R., \& Tully, R.~B.\ 2002, \apj, 580, 774

\bibitem[Ezoe et al.(2006)]{Ezoe06} Ezoe, Y., Kokubun, M., Makishima, K., Sekimoto, Y., \& Matsuzaki, K.\ 2006, \apj, 638, 860

\bibitem[Feast(1999)]{Feast99} Feast, M.\ 1999, New Views of the Magellanic Clouds, 190, 542

\bibitem[Feigelson et al.(2005)]{Feigelson05} Feigelson, E.~D., et al.\ 2005, \apjs, 160, 379 

\bibitem[Feigelson et al.(2011)]{Feigelson11} Feigelson, E.~D., et al.\ 2011, \apjs, submitted (CCCP Clustering Paper)

\bibitem[Figer(2008)]{Figer08} Figer, D.~F.\ 2008, IAU Symposium, 250, 247 

\bibitem[Figuer{\^e}do et al.(2002)]{Figueredo02} Figuer{\^e}do, E., Blum, R.~D., Damineli, A., \& Conti, P.~S.\ 2002, \aj, 124, 2739

\bibitem[Garcia(1994)]{Garcia94} Garcia, B.\ 1994, \apj, 436, 705

\bibitem[Garmire et al.(2003)]{Garmire03} Garmire, G.~P., Bautz, M.~W., Ford, P.~G., Nousek, J.~A., \& Ricker, G.~R., Jr.\ 2003, \procspie, 4851, 28

\bibitem[Getman et al.(2005)]{Getman05} Getman, K.~V., et al.\ 2005, \apjs, 160, 319

\bibitem[Georgelin et al.(2000)]{Georgelin00} Georgelin, Y.~M., Russeil, D., Amram, P., Georgelin, Y.~P., Marcelin, M., Parker, Q.~A., \& Viale, A.\ 2000, \aap, 357, 308

\bibitem[Girardi et al.(1997)]{Girardi97} Girardi, L., Bica, E., Pastoriza, M.~G., \& Winge, C.\ 1997, \apj, 486, 847 

\bibitem[G{\"u}del et al.(2008)]{Gudel08} G{\"u}del, M., Briggs, K.~R., Montmerle, T., Audard, M., Rebull, L., \& Skinner, S.~L.\ 2008, Science, 319, 309 

\bibitem[Hamaguchi et al.(2009)]{Hamaguchi09} Hamaguchi, K., et al.\ 2009, \apjl, 695, L4 

\bibitem[Hanson et al.(1997)]{Hanson97} Hanson, M.~M., Howarth, I.~D., \& Conti, P.~S.\ 1997, \apj, 489, 698 


\bibitem[Hickox \& Markevitch(2006)]{Hickox06} Hickox, R.~C., \& Markevitch, M.\ 2006, \apj, 645, 95

\bibitem[Hoffmeister et al.(2008)]{Hoffmeister08} Hoffmeister, V.~H., Chini, R., Scheyda, C.~M., Schulze, D., Watermann, R., N{\"u}rnberger, D., \& Vogt, N.\ 2008, \apj, 686, 310

\bibitem[Humphreys(1978)]{Humphreys78} Humphreys, R.~M.\ 1978, \apjs, 38, 309

\bibitem[Kuhn et al.(2010)]{Kuhn10} Kuhn, M.~A., Getman, K.~V., Feigelson, E.~D., Reipurth, B., Rodney, S.~A., \& Garmire, G.~P.\ 2010, \apj, 725, 2485

\bibitem[Lallement(2004)]{Lallement04} Lallement, R.\ 2004, \aap, 422, 391

\bibitem[Maercker et al.(2006)]{Maercker06} Maercker, M., Burton, M.~G., \& Wright, C.~M.\ 2006, \aap, 450, 253

\bibitem[Manchester et al.(2001)]{Manchester01} Manchester, R.~N., et al.\ 2001, \mnras, 328, 17 

\bibitem[Massey \& Hunter(1998)]{Massey98} Massey, P., \& Hunter, D.~A.\ 1998, \apj, 493, 180 

\bibitem[Moffat et al.(2002)]{Moffat02} Moffat, A.~F.~J., et al.\ 2002, \apj, 573, 191

\bibitem[Muller et al.(1998)]{Muller98} Muller, G.~P., Reed, R., Armandroff, T., Boroson, T.~A., \& Jacoby, G.~H.\ 1998, \procspie, 3355, 577

\bibitem[Muno et al.(2006)]{Muno06} Muno, M.~P., Law, C., Clark, J.~S., Dougherty, S.~M., de Grijs, R., Portegies Zwart, S., \& Yusef-Zadeh, F.\ 2006, \apj, 650, 203 

\bibitem[Ogura \& Ishida(1976)]{Ogura76} Ogura, K., \& Ishida, K.\ 1976, \pasj, 28, 35 

\bibitem[Parker et al.(2005)]{Parker05} Parker, Q.~A., et al.\ 2005, \mnras, 362, 689 

\bibitem[Persi et al.(1994)]{Persi94} Persi, P., Roth, M., Tapia, M., Ferrari-Toniolo, M., \& Marenzi, A.~R.\ 1994, \aap, 282, 474

\bibitem[Povich et al.(2007)]{Povich07} Povich, M.~S., et al.\ 2007, \apj, 660, 346 

\bibitem[Povich et al.(2009)]{Povich09} Povich, M.~S., et al.\ 2009, \apj, 696, 1278

\bibitem[Preibisch et al.(2005)]{Preibisch05} Preibisch, T., et al.\ 2005, \apjs, 160, 401

\bibitem[Price et al.(2001)]{Price01} Price, S.~D., Egan, M.~P., Carey, S.~J., Mizuno, D.~R., \& Kuchar, T.~A.\ 2001, \aj, 121, 2819 

\bibitem[Ranalli et al.(2008)]{Ranalli08} Ranalli, P., Comastri, A., Origlia, L., \& Maiolino, R.\ 2008, \mnras, 386, 1464

\bibitem[Rochau et al.(2010)]{Rochau10} Rochau, B., Brandner, W., Stolte, A., Gennaro, M., Gouliermis, D., Da Rio, N., Dzyurkevich, N., \& Henning, T.\ 2010, \apjl, 716, L90

\bibitem[Russell \& Dopita(1990)]{Russell90} Russell, S.~C., \& Dopita, M.~A.\ 1990, \apjs, 74, 93

\bibitem[Sana et al.(2010)]{Sana10} Sana, H., Momany, Y., Gieles, M., Carraro, G., Beletsky, Y., Ivanov, V.~D., de Silva, G., \& James, G.\ 2010, \aap, 515, A26

\bibitem[Smith(2006)]{Smith06} Smith, N.\ 2006, \apj, 644, 1151 

\bibitem[Smith \& Brooks(2008)]{Smith08} Smith, N., \& Brooks, K.~J.\ 2008, Handbook of Star Forming Regions, Volume II, ed.\ B.\ Reipurth (San Francisco, CA:  ASP), 138

\bibitem[Smith et al.(2010)]{Smith10} Smith, N., Bally, J., \& Walborn, N.~R.\ 2010, \mnras, 405, 1153 

\bibitem[Tapia et al.(2003)]{Tapia03} Tapia, M., Roth, M., V{\'a}zquez, R.~A., \& Feinstein, A.\ 2003, \mnras, 339, 44 

\bibitem[Townsley et al.(2003)]{Townsley03} Townsley, L.~K., Feigelson, E.~D., Montmerle, T., Broos, P.~S., Chu, Y.-H., \& Garmire, G.~P.\ 2003, \apj, 593, 874 

\bibitem[Townsley et al.(2006a)]{Townsley06a} Townsley, L.~K., Broos, P.~S., Feigelson, E.~D., Brandl, B.~R., Chu, Y.-H., Garmire, G.~P., \& Pavlov, G.~G.\ 2006a, \aj, 131, 2140

\bibitem[Townsley et al.(2006b)]{Townsley06b} Townsley, L.~K., Broos, P.~S., Feigelson, E.~D., Garmire, G.~P., \& Getman, K.~V.\ 2006b, \aj, 131, 2164 

\bibitem[Townsley(2009a)]{Townsley09a} Townsley, L.~K.\ 2009a, AIP Conf.\ Proc.\ 1156, ``The Local Bubble and Beyond II,'' ed. K.~D.~Kuntz, R.~K.~Smith, \& S.~L.~Snowden, p.\ 225

\bibitem[Townsley(2009b)]{Townsley09b} Townsley, L.~K.\ 2009b, STScI Symp.\ Series 20, ``Massive Stars:  From Pop III and GRBs to the Milky Way,'' ed.\ M.~Livio \& E.~Villaver, p.\ 60 (arXiv:astro-ph/0608173)

\bibitem[Townsley et al.(2011a)]{Townsley11a} Townsley, L.~K., et al.\ 2011a, \apjs, submitted (CCCP Intro Paper)

\bibitem[Townsley et al.(2011b)]{Townsley11b} Townsley, L.~K., et al.\ 2011b, \apjs, submitted (CCCP Diffuse Paper)

\bibitem[Wang \& Helfand(1991)]{Wang91} Wang, Q., \& Helfand, D.~J.\ 1991, \apj, 370, 541

\bibitem[Wang et al.(2011)]{Wang11} Wang, J., et al.\ 2011, \apjs, submitted (CCCP Tr15 Paper)


\bibitem[Wolk et al.(2006)]{Wolk06} Wolk, S.~J., Spitzbart, B.~D., Bourke, T.~L., \& Alves, J.\ 2006, \aj, 132, 1100

\end{thebibliography}
\end{document}